\documentclass[aps,prb,superscriptaddress,amsmath,twocolumn]{revtex4-2}

\usepackage{amsmath,amsfonts,amssymb,mathtools,braket,graphicx,enumitem}
\usepackage[utf8]{inputenc}
\usepackage{layouts}
\usepackage[colorlinks=true,citecolor=blue,linkcolor=red,urlcolor=blue]{hyperref}
\usepackage[normalem]{ulem}
\usepackage[percent]{overpic}

\newcommand{\euler}{\mathrm{e}}
\newcommand{\iu}{\mathrm{i}}

\begin{document}

\title{Differential scattering cross section of the non-Abelian Aharonov--Bohm effect in multiband systems}

\author{Róbert Németh}
\author{ J{\'o}zsef Cserti}
\affiliation{Department of Physics of Complex Systems,
	ELTE E{\"o}tv{\"o}s Lor{\'a}nd University,
	H-1117 Budapest, P\'azm\'any P{\'e}ter s{\'e}t\'any 1/A, Hungary
}

\begin{abstract}
We develop a unified treatment of the non-Abelian Aharonov--Bohm (AB) effect in isotropic multiband systems, namely, the scattering of particles on a gauge field corresponding to a non-commutative Lie group.
We present a complex contour integral representation of the scattering states for such systems, and, using their asymptotic form, we calculate the differential scattering cross section. The angular dependence of the cross section turns out to be the same as that obtained originally by Aharonov and Bohm in their seminal paper, but this time it depends on the polarization of the incoming plane wave. 
As an application of our theory, we perform the contour integrals for the wave functions explicitly and calculate the corresponding cross section for three non-trivial isotropic multiband systems relevant to condensed matter and particle physics. To have a deeper insight into the nature of the scattering, we plot the probability and current distributions for different incoming waves. 
This paper is a generalization of our recent results on the Abelian AB effect providing an extension of exactly solvable AB scattering problems. 
\end{abstract}

\maketitle

\section{Introduction}
\label{sec1}

Aharonov and Bohm in their seminal paper~\cite{Aharonov-Bohm:cikk} calculated how the incident plane wave of a spinless free particle is scattered by an infinitely thin magnetic solenoid and the differential scattering cross section. The experimental verification of this quantum effect was first reported by Chambers~\cite{PhysRevLett.5.3}, while the mathematical details of the theory were further explored by Berry~\cite{Berry_1980,Berry_Chambers_1980}. The scattering problem was also generalized in the framework of relativistic quantum mechanics by Alford and Wilczek~\cite{Wilczek_PhysRevLett.62.1071}, by Gerbert~\cite{Gerbert_PhysRevD.40.1346}, and also by Hagen and Ramaswamy~\cite{Hagen_PhysRevLett.64.503, Hagen_PhysRevD.42.3524}. The non-Abelian version of the Aharonov--Bohm effect in which the magnetic field responsible for the scattering is replaced by an $\mathrm{SU}(2)$ gauge field has been  studied by Wu and Yang~\cite{Wu:NA}, and by Horváthy~\cite{Horvathy:NA}. 
They discussed the analogies of both the original Aharonov--Bohm scattering problem and the interference pattern forming in two-slit geometries such as in the experiment of Chambers. Lately, the theory was also extended to the case of time-dependent gauge fields~\cite{Bright:2015rsa,HosseiniMansoori:2016rex}. However, as far as we know, only a limited number of specific systems were discussed, and a general formalism was never developed. 

Recently, the realization of non-Abelian gauge potentials in optical lattices was studied extensively~\cite{Goldman:paper1,Bermudez:paper,Gorshkov:2010Nature,Barnett:paper,PhysRevLett.110.125303,Goldman:paper2,Tagliacozzo:2012df,PhysRevLett.110.125304}, with special interest towards the non-Abelian AB effect~\cite{Osterloh:paper,Jacob:2007jb,Dalibard:2010ph,Goldman_2014,Huo:paper,AIDELSBURGER2018394}. Additionally, Yang \emph{et al.\ }reported the first successful observation of the SU(2) Aharonov--Bohm phase shift using optical interferometry~\cite{Yang:2019Science}. 
Therefore, the role of gauge fields is becoming more and more significant in the current research of condensed matter physics.

In our recent work~\cite{Nemeth-Cserti:paper}, based on a contour integral formalism, we developed a general theory of the Abelian Aharonov--Bohm effect for multiband systems possessing isotropic dispersion relations to obtain the scattering states and the differential cross section. 
As a striking result, it turned out that the cross section has the same functional form for \emph{all isotropic systems} and is identical to that derived by Aharonov and Bohm~\cite{Aharonov-Bohm:cikk}. 
In this work, as an extension, we develop a general theory for the case of non-Abelian scattering potentials.  
In particular, we find that the cross section again has the same angular dependence as in the Abelian case and is given by 
\begin{equation}
	\sigma_\mathrm{AB,gen}(\varphi) = \frac{\Sigma}{2\pi k \cos^2(\varphi/2)},
	\label{eqGen0}
\end{equation} 
where $\Sigma$ is a numerical factor depending on the non-Abelian nature of the flux and the incoming wave function. The detailed derivation of this result is presented in the following sections.
To see how our approach can be used in practice, we shall consider three simple examples relevant 
to both solid-state and particle physics. However, more complicated multiband systems can be treated in a similar fashion.

The paper is structured as follows. In Sec.~\ref{sec2}, we present the general results of our theory. Specifically, in Sec.~\ref{sec21}, we introduce the basic notions of gauge theory needed to define the scattering problem; in Sec.~\ref{sec22}, we describe the set of systems our approach is applicable to; in Sec.~\ref{sec23}, we give a general formula for the scattering states using an integral representation; and in Sec.~\ref{sec24}, we derive the differential scattering cross section and discuss similarities and differences compared to the Abelian case. In Sec.~\ref{sec3}, our method is applied to one- and two-band systems and explicit expressions of both the scattering states and the differential cross section are presented. In particular, we consider the case of an SU(2) doublet in Sec.~\ref{sec31}, an SU(3) triplet in Sec.~\ref{sec32}, and a U(2) doublet in Sec.~\ref{sec33}.

\section{General theory}
\label{sec2}

In this section, we outline our approach to obtain the scattering states and the cross section for the non-Abelian AB effect. First, we define the most general Hamiltonian describing isotropic multiband systems. Then, we propose our contour integral representation of the scattering states which is a generalization of that presented in Ref.~\onlinecite{Nemeth-Cserti:paper}. Furthermore, we calculate the differential scattering cross section.

\subsection{Gauge potential}
\label{sec21}

In the conventional Aharonov--Bohm effect, the physical space is taken to be the two-dimensional plane with the origin removed; we denote this by $\mathbb{R}^2_\circ$. The scattering potential is introduced in the form of the magnetic vector potential which is a (co-)vector field on the physical space ~\cite{Aharonov-Bohm:cikk,Nemeth-Cserti:paper}. Each component of this field is real, so the target space is the set $\mathbb{R}$ which can be identified with the one-dimensional Lie algebra $\mathfrak{u}(1)$ corresponding to the Lie group $\mathrm{U}(1)$ of complex numbers with absolute value one. In this way, one can think of the vector potential as a $\mathrm{U}(1)$ gauge potential. 

This description allows us to create the concept of \emph{gauge potentials} corresponding to a more general non-Abelian Lie group $G$. For our investigation, we restrict ourselves to groups that are \emph{locally isomorphic to a subgroup of $\mathrm{U}(N)$ around the identity element}. In this case, each component of the gauge potential is an element of the Lie algebra $\mathfrak{g}\subseteq\mathfrak{u}(N)$, that is, a self-adjoint $N\times N$ complex matrix. Along these lines, the generalization of the AB vector potential to the $G$-case is straightforward, in Cartesian components given by
\begin{subequations}
	\label{eqGen1}
	\begin{align}
		\hat{A}_x(x,y) &= -\frac{\hat{\Phi} y}{2\pi(x^2 + y^2)} , \label{eqGen1a} \\
		\hat{A}_y(x,y) &= \frac{\hat{\Phi} x}{2\pi(x^2 + y^2)} . 
		\label{eqGen1b}
	\end{align}
\end{subequations} Here the non-Abelian version of the flux $\hat{\Phi}$ is an element of $\mathfrak{g}$. Self-adjointness implies that $\hat{\Phi}$ is diagonal on a suitable orthonormal basis and its eigenvalues $\Phi_n$ are real for all $n\in\{1,...,N\}$.

The \emph{gauge field} is an antisymmetric rank-2 tensor field that serves as the generalization of the magnetic field. Thus, in two dimensions, it has a single independent component given by
\begin{equation}
	\hat{F}_{xy} = \partial_x\hat{A}_y - \partial_y\hat{A}_x + \big[\hat{A}_x,\hat{A}_y\big] = 0 .
	\label{eqGen4}
\end{equation} Similarly as before, this field vanishes on $\mathbb{R}^2_\circ$. 

For the Abelian case, the line integral of the gauge potential equals the flux for an arbitrary closed curve $C$ once encircling the origin anticlockwise. For the non-Abelian case, however, we should instead use the path-ordered exponential of the gauge potential called the \emph{Wilson loop operator}:
\begin{equation}
	\hat{W} = \mathrm{P}\exp\left[\frac{\iu e}{\hbar}\oint_C \boldsymbol{\hat{A}}\right] = \exp\left(2\pi\iu\hat{\alpha}\right) ,
	\label{eqGen5}
\end{equation} where we introduced the dimensionless flux $\hat{\alpha} = \hat{\Phi}/\Phi_0$ with $\Phi_0 = h/e$. 
In summary, the infinitesimal flux tube has similar properties in the Abelian and non-Abelian cases, but for the latter the matrix nature of the flux must be kept in mind.

\subsection{Hamiltonian operator}
\label{sec22}

The most general Hilbert space of the systems we study in this work takes 
the form $\mathcal{H} = L^2(\mathbb{R}^2)\otimes\mathbb{C}^N\otimes\mathbb{C}^D$ consisting of the two-dimensional spatial, $N$-dimensional gauge, and a $D$-dimensional internal degrees of freedom. The gauge degree of freedom is also referred to as \emph{polarization}. Furthermore, we limit ourselves to studying Hamiltonians satisfying the three requirements mentioned in our previous paper~\cite{Nemeth-Cserti:paper}: \emph{polynomicity}, \emph{isotropy}, and \emph{regularity}. Thus, the Hamiltonian operator in the absence of the gauge field takes the form
\begin{equation}
	\hat{H} = \sum_{i = 0}^I \sum_{j = 0}^J \hat{p}_x^{i} \hat{p}_y^{j} \otimes\hat{I}_{N}\otimes\hat{T}_{ij} ,
	\label{eqGen6}
\end{equation} where $\hat{I}_N$ is the $N\times N$ identity matrix. In the presence of the gauge field, we need to introduce the kinetic momentum operators acting on the direct product Hilbert space $L^2(\mathbb{R}^2)\otimes\mathbb{C}^N$:
\begin{subequations}
	\label{eqGen7}
	\begin{align}
		\hat{\Pi}_x = \hat{p}_x\otimes\hat{I}_{N} + \hat{A}_x(\hat{x},\hat{y}) , \label{eqGen7a} \\ 
		\hat{\Pi}_y = \hat{p}_y\otimes\hat{I}_{N} + \hat{A}_y(\hat{x},\hat{y}) . 
		\label{eqGen7b}
	\end{align}
\end{subequations} Here the components of the gauge potential from Eq.~\eqref{eqGen1} are used. Then the total Hamiltonian in the presence of the gauge field reads
\begin{equation}
	\hat{H} = \sum_{i = 0}^I \sum_{j = 0}^J \hat{\Pi}_x^{i} \hat{\Pi}_y^{j} \otimes\hat{T}_{ij} .
	\label{eqGen10}
\end{equation} 

\subsection{Scattering states}
\label{sec23}

As in our previous work~\cite{Nemeth-Cserti:paper}, here we construct the contour integral representation of the scattering states for the non-Abelian AB effect. 
To this end, we need the eigenvectors of the Hamiltonian \eqref{eqGen10} in the absence of the gauge field. These are plane waves with wave number $\boldsymbol{k} = k(\cos\vartheta,\sin \vartheta)$, definite polarization $\mathbf{w}_n\in\mathbb{C}^N$ and internal degree of freedom $\mathbf{u}_s(\boldsymbol{k})\in\mathbb{C}^D$ given by:
\begin{equation}
	\boldsymbol{\Psi}_{s,n,\boldsymbol{k}}(r,\varphi) = \euler^{\iu kr\cos(\varphi - \vartheta)} \mathbf{w}_n \mathbf{u}_s(\boldsymbol{k}) .
	\label{eqGen11}
\end{equation} 
The corresponding eigenvalues $E_s(k)$ consists of $N$-fold degenerate bands owing to the different polarizations. Without the loss of generality, we can choose the vectors $\mathbf{w}_n$ such that they are eigenvectors of $\hat{\alpha}$ with eigenvalues $\alpha_n = \Phi_n/\Phi_0$:
\begin{equation}
	\hat{\alpha}\mathbf{w}_n = \alpha_n\mathbf{w}_n .
	\label{eqGen12}
\end{equation} This implies that they are eigenvalues of the Wilson loop operator $\hat{W}$ as well: 
\begin{equation}
	\hat{W}\mathbf{w}_n = \euler^{2\pi\iu\alpha_n}\mathbf{w}_n .
	\label{eqGen13}
\end{equation}

Now, the scattering states satisfying the necessary requirements (i)-(iii) detailed in our paper~\cite{Nemeth-Cserti:paper} are given by
\begin{equation}
	\begin{aligned}
		\boldsymbol{\Psi}&^{(+)}_{s,n,\boldsymbol{k}}(r,\varphi) = \sum_{m = -\infty}^{\infty} \frac{\epsilon(m+\alpha_n)}{2\pi} \\
		&\times\int
		\displaylimits^{\phantom{\Gamma}}_{\Gamma(m+\alpha_n,\varphi)}\mathrm{d}\xi~ \boldsymbol{\Psi}_{s,n,\boldsymbol{K}}(r,\varphi) \mathrm{e}^{\iu m(\xi - \vartheta) - \iu\alpha_n(\varphi - \xi)} ,
		\label{eqGen14}
	\end{aligned}
\end{equation} where $\epsilon(x)$ is the sign function defined as $\epsilon(x) =1$, if $x\ge 0$ and $\epsilon(x) =-1$, if $x< 0$. Furthermore, $\boldsymbol{K} = k {\left[\cos \xi, \sin \xi\right]}$ with $\xi$ being a complex integration variable. The integration contours $\Gamma(m+\alpha_n,\varphi)$ are curves on the complex plane depending on the sign of $m+\alpha_n$ and the value of the real space polar angle $\varphi$:
\begin{equation}
	\Gamma(m+\alpha_n,\varphi) = \begin{cases} \Gamma_+(\varphi),~\mathrm{if}~m+\alpha_n\ge 0, \\ \Gamma_-(\varphi),~\mathrm{if}~m+\alpha_n< 0.  \end{cases}
	\label{contours:eq}
\end{equation}
The curves $\Gamma_+(\varphi)$ and $\Gamma_-(\varphi)$ further depend on the sign of the radial component $v_{s,k}$ of the group velocity 
\begin{equation}
	\boldsymbol{v}_{s} = \frac{1}{\hbar} \frac{\partial E_{s}}{\partial \boldsymbol{k}}.
	\label{group_velocity:eq}
\end{equation} 

In particular, if $v_{s,k}>0$, the curve $\Gamma_+(\varphi)$ is $\cup$-shaped running from $\xi = -5\pi/2 + \varphi + \iu\infty$ to $\xi = -\pi/2 + \varphi + \iu\infty$ with $\mathrm{Re}(\xi)>0$, and the curve $\Gamma_-(\varphi)$ is $\cap$-shaped running from $\xi = -3\pi/2 + \varphi - \iu\infty$ to $\xi = \pi/2 + \varphi - \iu\infty$ with $\mathrm{Re}(\xi)<0$. However, if $v_{s,k}<0$, the curve $\Gamma_+(\varphi)$ must be shifted by $2\pi$ along the real axis compared to the previous definition. Furthermore, if $v_{s,k}=0$, the scattering state has no physical meaning as the corresponding plane waves have constant zero current density indicating that they do not propagate.

By a straightforward generalization of the proofs given in our previous paper~\cite{Nemeth-Cserti:paper}, one can show that the wave functions given by Eq.~(\ref{eqGen14}) are indeed proper scattering states satisfying the necessary requirements (i)-(iii) detailed there.
Note that using the eigenvector basis given by Eq.~\eqref{eqGen12}, the non-Abelian scattering problem is decomposed into $N$ different Abelian counterparts. 
Then an arbitrary incoming wave can be given as a linear combination of the wave functions~\eqref{eqGen11}:
\begin{equation}
	\boldsymbol{\Psi}_\mathrm{gen} = \sum_{n=1}^N c_n \boldsymbol{\Psi}_{s,n,\boldsymbol{k}} ,
	\label{eqGen15}
\end{equation} 
hence the same linear combination of the scattering states of Eq.~\eqref{eqGen14} can be used: 
\begin{equation}
	\boldsymbol{\Psi}_\mathrm{gen}^{(+)} = \sum_{n=1}^N c_n \boldsymbol{\Psi}_{s,n,\boldsymbol{k}}^{(+)} .
	\label{eqGen16}
\end{equation} 

\subsection{Differential scattering cross section}
\label{sec24}

As a usual procedure~\cite{Nemeth-Cserti:paper}, using the asymptotic form of the scattering states~\eqref{eqGen14} 
we now calculate the differential cross section.  
Here we assume that the incident particles are coming from the direction of the positive $x$ axis. 
Then for given flux parameters $\alpha_n$ we find that 
the differential cross section is:
\begin{equation}
	\sigma_{\mathrm{AB},n}(\varphi) \equiv \frac{\mathrm{d}\sigma}{\mathrm{d}\varphi}(s,n,k;\varphi) = \frac{\sin^2(\alpha_n\pi)}{2\pi k \cos^2(\varphi/2)} \;.
	\label{eqGen17}
\end{equation} 
For an arbitrary incoming wave, the different terms in Eq.~\eqref{eqGen16} are pairwise orthogonal due to the corresponding polarization vectors $\mathbf{w}_n$. Therefore, the cross section becomes a weighted sum of the terms in Eq.~\eqref{eqGen17} as 
\begin{equation}
	\sigma_\mathrm{AB,gen}(\varphi) \equiv \frac{\mathrm{d}\sigma}{\mathrm{d}\varphi}(s,k;\varphi) = \frac{\Sigma}{2\pi k \cos^2(\varphi/2)} ,
	\label{eqGen18}
\end{equation} where we introduced the dimensionless \emph{cross section factor} $\Sigma$ given by
\begin{equation}
	\label{eqGen18_b}
	\Sigma = \sum_{n=1}^N |c_n|^2 \, \sin^2(\alpha_n\pi) .
\end{equation} 
As an important corollary, these results again show that the differential cross section is independent of the specific form of the dispersion relation $E_s(k)$ provided that it is isotropic. Note that when the angular distribution and the polarization are measured simultaneously then the cross section will be different from that given in Eq.~\eqref{eqGen18}, but can easily be calculated by properly taking into account the current contributions. However, the details of this issue are not considered in this paper.

\section{Applications}
\label{sec3}

In this section, we present three examples that provide more insight into the nature of the non-Abelian AB scattering problem. These examples are relevant in both condensed matter and particle physics. One of them has already been studied before (see Sec.~\ref{sec31}) but, as far as we know, the other two are new (see Secs.~\ref{sec32} and~\ref{sec33}).

\subsection{SU(2) doublet}
\label{sec31}

The first example is the \emph{SU(2) doublet}. Such a system can appear in the non-relativistic model of nucleons~\cite{Heisenberg:1932} where protons and neutrons are regarded as different isospin eigenstates of the same particle, analogously to a spin-$1/2$ degree of freedom~\cite{Cheng:1984vwu,Nachtmann:1990ta,Horvath:2019atn}. The non-Abelian Aharonov--Bohm effect in this system has been studied before by Wu and Yang~\cite{Wu:NA}, and also by Horváthy~\cite{Horvathy:NA}.
Note that such a system is also relevant in condensed matter physics, as recently similar systems were synthesized by trapping ultracold atoms in optical square lattices~\cite{Goldman:paper2,Huo:paper}. The low energy dynamics of the atomic motion mimic that of two-dimensional non-relativistic fermions. In that case, the internal degree of freedom corresponding to the isospin emerges as doubly split energy eigenstates of the cold atoms, and the gauge field is generated for instance by laser-assisted tunneling~\cite{Goldman_2014,AIDELSBURGER2018394}.

Using our methods developed in Sec.~\ref{sec2}, we now reproduce the results obtained by Horváthy~\cite{Horvathy:NA}.
The Hilbert space corresponding to the SU(2) doublet is $\mathcal{H} = L^2(\mathbb{R}^2,\mathbb{C})\otimes\mathbb{C}^2$, that is, there is no internal degree of freedom ($D=1$), only a two-dimensional gauge degree of freedom ($N=2$) corresponding to polarization. The Hamiltonian operator $\hat{H}: \mathcal{D}_H \to \mathcal{H}$ is a quadratic polynomial of the momentum operators:
\begin{equation}
	\hat{H} = \frac{1}{2M} \left(\hat{p}_x^2 + \hat{p}_y^2\right)\otimes\hat{I}_2 ,
	\label{eqIso1}
\end{equation} where $M$ is the mass of the particles, and $\hat{I}_2$ is the $2\times 2$ identity matrix. 

For wave number $\boldsymbol{k} = k(\cos\vartheta,\sin \vartheta)$, the energy eigenvalues of the Hamiltonian in Eq.~\eqref{eqIso1} are doubly degenerate, and are given by
\begin{equation}
	E(k,\vartheta) = \frac{\hbar^2 k^2}{2M} .
	\label{eqIso4}
\end{equation} 
One can see that the band structure is isotropic, namely, depends only on the magnitude of wave number $k = |\boldsymbol{k}|$, and thus the constant energy curves are circles.

Due to the regularity requirement~\cite{Nemeth-Cserti:paper}, the momentum space eigenvectors need to be chosen such that one of its components is independent of $\vartheta$. In this case, there is only one possible choice up to a constant multiplier:
\begin{equation}
	\mathbf{u}(k,\vartheta) = 1 .
	\label{eqIso6}
\end{equation}

From Eqs.~\eqref{group_velocity:eq} and~\eqref{eqIso4}, one finds that the group velocity is
\begin{align}
	v_k(k,\vartheta) = \frac{\hbar k}{M},    
	\quad  v_\vartheta(k,\vartheta) = 0 . 
	\label{eqIso7b}
\end{align} 
The radial component of the group velocity is positive for all $k>0$ implying that the single band of this model is an electron-like band. Here we stress that the sign of the radial group velocity is relevant because it modifies the contours on the complex plane as detailed after Eq.~\eqref{group_velocity:eq}. 

The $\mathfrak{su}(2)$ Lie algebra corresponding to the SU(2) group is three-dimensional, whose standard basis consists of the half-Pauli matrices:
\begin{equation}
\begin{aligned}
	\hat{\tau}_1 &= \frac12 \begin{pmatrix}
		0 & 1 \\ 1 & 0
	\end{pmatrix},\quad \hat{\tau}_2 = \frac12 \begin{pmatrix}
		0 & -\iu \\ \iu & 0
	\end{pmatrix},\\ \hat{\tau}_3 &= \frac12 \begin{pmatrix}
		1 & 0 \\ 0 & -1
	\end{pmatrix}.
	\label{eqIso8}
\end{aligned}
\end{equation} The gauge potential of this problem is given according to Eq.~\eqref{eqGen1} where the SU(2)-flux $\hat{\Phi}$, or equivalently the dimensionless flux $\hat{\alpha}$ can be parameterized with a real parameter $\alpha$ as follows
\begin{equation}
	\hat{\alpha} = 2\alpha\hat{\tau}_3 = \begin{pmatrix}
		\alpha & 0 \\ 0 & -\alpha
	\end{pmatrix} .
	\label{eqIso9}
\end{equation} 
Note that owing to the self-adjointness of $\mathfrak{su}(2)$-generators, 
an arbitrary flux can be diagonalized as in \eqref{eqIso9}, 
and a similar statement holds also for the group elements, as pointed out by Horváthy ~\cite{Horvathy:NA}.
The eigenvectors of $\hat{\alpha}$ take the form
\begin{equation}
	\mathbf{w}_1 = \begin{pmatrix}
		1 \\ 0
	\end{pmatrix} ,\quad \mathbf{w}_2 = \begin{pmatrix}
		0 \\ 1
	\end{pmatrix} ,
	\label{eqIso10}
\end{equation} with eigenvalues $\alpha$ and $-\alpha$, respectively.

We now introduce the probability density of the position corresponding to an arbitrary state $(\Psi_1,\Psi_2)\in\mathcal{H}$: 
\begin{equation}
	\varrho = \Psi_1^* \Psi_1 + \Psi_2^* \Psi_2 ,
	\label{eqIso11}
\end{equation} 
and calculate the probability current density using the general formula given in our previous work~\cite{Nemeth-Cserti:paper}. Then, we find
\begin{subequations}
	\begin{align}
		j_x &= \frac{1}{M} \sum_{n=1}^2 \sum_{l=1}^2 \mathrm{Re}\left[\Psi_n^* \big(\hat{\Pi}_{x,nl} \Psi_l\big)\right] , \label{eqIso12a} \\
		j_y &= \frac{1}{M} \sum_{n=1}^2 \sum_{l=1}^2 \mathrm{Re}\left[\Psi_n^* \big(\hat{\Pi}_{y,nl} \Psi_l\big)\right] , 
		\label{eqIso12b}
	\end{align}
\end{subequations} where $\hat{\Pi}_{x,nl}$ and $\hat{\Pi}_{y,nl}$ are the matrix elements of the kinetic momentum operators.

We now apply Eq.~\eqref{eqGen14} to calculate the scattering states. The integral that appears in this expression can be simplified using the Schäfli--Sommerfeld integral formula of the Bessel functions of the first kind ~\cite{HTF:book}, and the final result becomes
\begin{subequations}
	\label{eqIso13-14_egybe}
\begin{equation}
	\begin{aligned}
		\boldsymbol{\Psi}^{(+)}_{1,\boldsymbol{k}}(r,\varphi) = \sum_{m=-\infty}^\infty &(-\iu)^{|m+\alpha|} J_{|m+\alpha|}(k r) \\ 
		&\times\euler^{\iu m(\varphi - \vartheta + \pi)} \begin{pmatrix} 1 \\ 0 \end{pmatrix} ,
	\end{aligned}
	\label{eqIso13}
\end{equation} 
for $n=1$, and 
\begin{equation}
	\begin{aligned}
		\boldsymbol{\Psi}^{(+)}_{2,\boldsymbol{k}}(r,\varphi) = \sum_{m=-\infty}^\infty &(-\iu)^{|m-\alpha|} J_{|m-\alpha|}(k r) \\
		&\times\euler^{\iu m(\varphi - \vartheta + \pi)} \begin{pmatrix} 0 \\ 1 \end{pmatrix} ,
	\end{aligned}
	\label{eqIso14}
\end{equation} 
\end{subequations}
for $n=2$. 
Note that for $\vartheta = 0$, these wave functions agree with the solutions obtained by Horváthy~\cite{Horvathy:NA}; thus, this calculation serves as a checkpoint for our general method presented in this work.

\begin{figure}[!hbt]
	\begin{overpic}[scale=0.5]{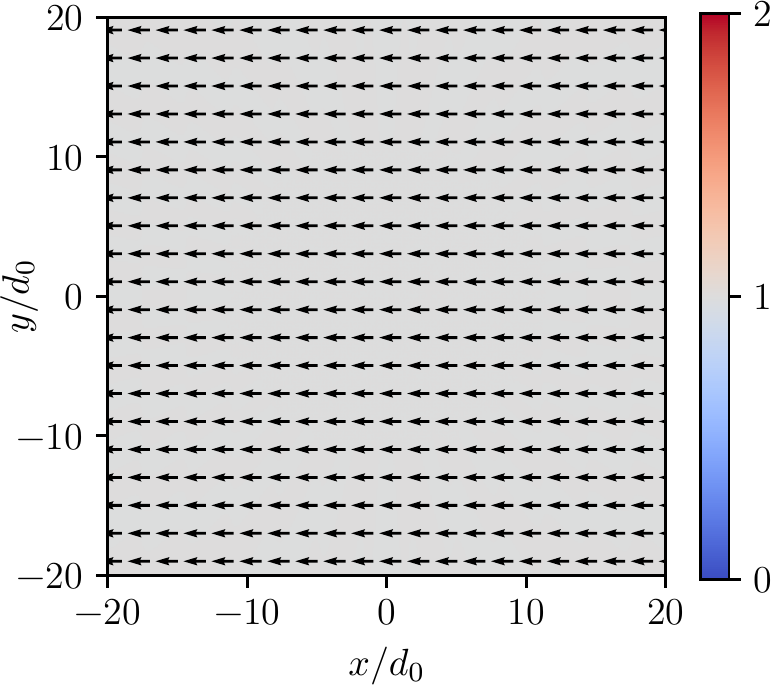}
		\put(-5,80){\scriptsize{(a)}}
	\end{overpic}
	\hspace{10pt}
	\begin{overpic}[scale=0.5]{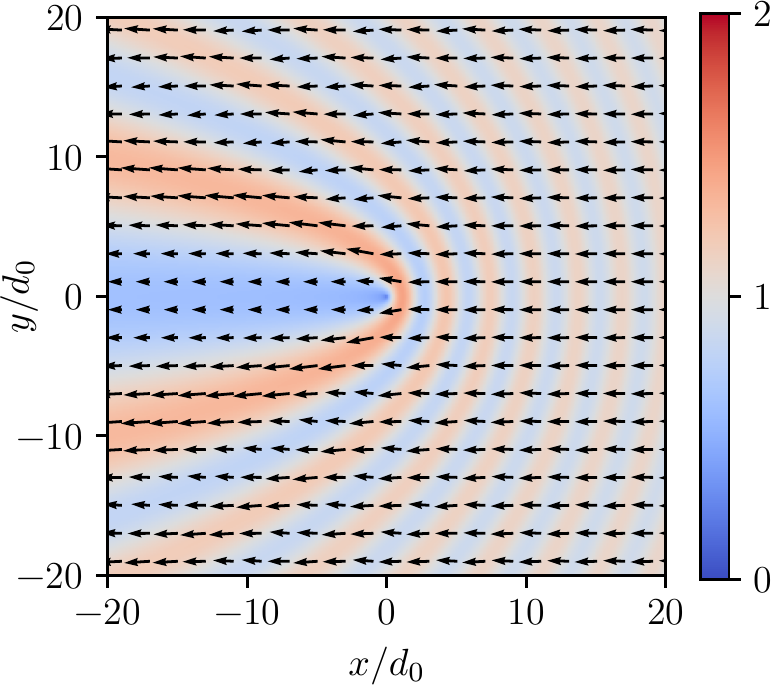}
		\put(-5,80){\scriptsize{(b)}}
	\end{overpic}
	\begin{overpic}[scale=0.5]{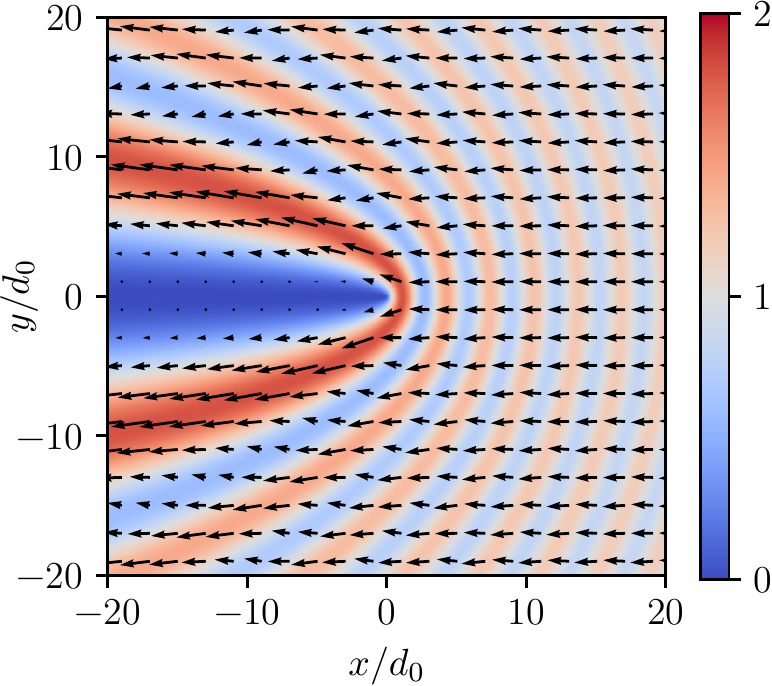}
		\put(-5,80){\scriptsize{(c)}}
	\end{overpic}
	\hspace{10pt}
	\begin{overpic}[scale=0.5]{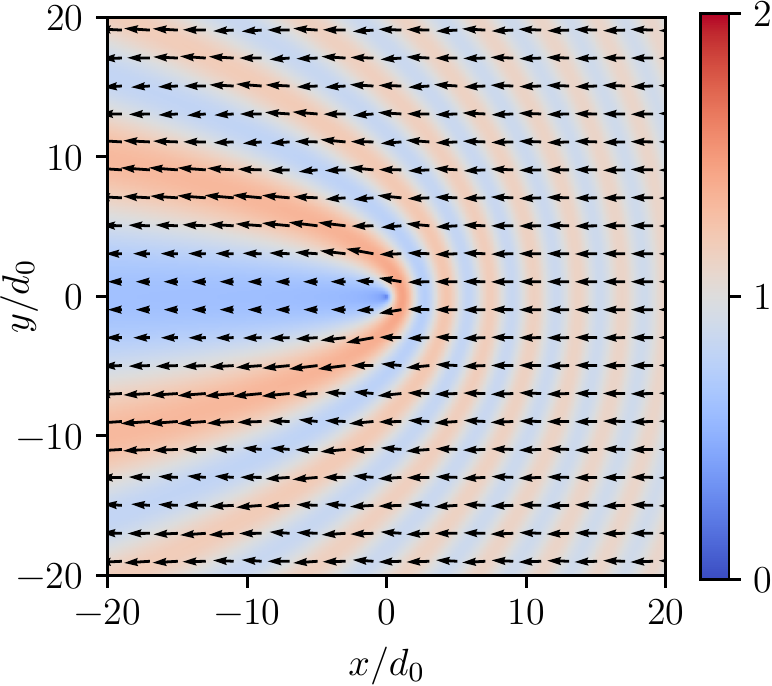}
		\put(-5,80){\scriptsize{(d)}}
	\end{overpic}
	\caption{Scattering states corresponding to the Aharonov--Bohm effect for the SU(2) doublet. The probability density $\varrho$ (colors) and current density $\boldsymbol{j}$ (arrows) are computed for $kd_0 = 1$ (where $d_0$ is a natural length unit) and (a) $\alpha = 0$, (b) $\alpha = 0.2$, (c) $\alpha = 0.5$, (d) $\alpha = 0.8$.}
	\label{figIso2}
\end{figure}

We now consider a special scattering state built as a linear combination of the wave functions in \eqref{eqIso13-14_egybe}:
\begin{equation}
	\boldsymbol{\Psi}^{(+)} = \frac{1}{\sqrt2}\left[\boldsymbol{\Psi}^{(+)}_{1,\boldsymbol{k}} + \boldsymbol{\Psi}^{(+)}_{2,\boldsymbol{k}}\right] .
	\label{eqIso15a}
\end{equation} 
To visualize this wave function, we calculate the probability density and the current density using Eqs.~\eqref{eqIso11}, \eqref{eqIso12a} and \eqref{eqIso12b}. The results of our numerical calculations are shown in Fig.~\ref{figIso2}.

As expected, they mostly resemble the plots obtained in the Abelian case (see e.g., Ref.~\onlinecite{Nemeth-Cserti:paper}) with some differences we point out now.
The most significant feature is that the scattering states are reflection-symmetric with respect to the $x$ axis. 
This is the consequence of the special choice of linear combination coefficients in Eq.~\eqref{eqIso15a}.
If this was not the case then such incidental symmetry would not be present. This is clearly seen, e.g., from the scattering states of either Eq.~\eqref{eqIso13} or \eqref{eqIso14}. These possess the same probability density and current density as their Abelian counterparts and are thereby asymmetric.

Finally, we calculate the cross section factor $\Sigma$ defined in Eq.~\eqref{eqGen18_b} for the SU(2) doublet, and obtain:
\begin{equation}
	\Sigma = \sin^2(\alpha\pi) ,
\end{equation}
which is independent of the linear combination coefficients $c_n$ used in Eq.~\eqref{eqGen15}. 
This shows that the differential cross-section is completely insensitive to the initial state, much like in the Abelian case. However, it must be noted that this is only due to the simple nature of the $\mathrm{SU}(2)$-group. In the following sections, we consider more complicated cases.

\subsection{SU(3) triplet}
\label{sec32}

Our second example is the \emph{SU(3) triplet}. This model is used to describe the non-relativistic dynamics of free quarks~\cite{GELLMANN1964214,Zweig:1964jf} which participate in strong interactions via a three-state degree of freedom coupling to the gluon field, called color ~\cite{Cheng:1984vwu,Nachtmann:1990ta,Horvath:2019atn}.
Although free quarks are rarely found in nature, an analogous system can be engineered using again ultracold atoms trapped in an optical lattice, and laser-induced transitions, but here, the energy eigenstates are triply degenerate~\cite{Barnett:paper,PhysRevLett.110.125303}. The non-Abelian AB effect in this system was proposed first by Horváthy~\cite{Horvathy:NA}, but to the best of our knowledge, the calculation of the scattering states was not performed before. In the following, we aim to do so using our method developed in Sec.~\ref{sec2}.

The Hilbert space corresponding to the SU(3) triplet is $\mathcal{H} = L^2(\mathbb{R}^2,\mathbb{C})\otimes\mathbb{C}^3$, that is, there is no internal degree of freedom ($D=1$), only a three-dimensional gauge degree of freedom ($N=3$) corresponding to polarization. The Hamiltonian operator $\hat{H}: \mathcal{D}_H \to \mathcal{H}$ is given as a quadratic polynomial of the momentum operators:
\begin{equation}
	\hat{H} = \frac{1}{2M} \left(\hat{p}_x^2 + \hat{p}_y^2\right)\otimes\hat{I}_3 ,
	\label{eqIso16}
\end{equation} where $M$ is the mass of the particles, and $\hat{I}_3$ is the $3\times 3$ identity matrix. The above expression is formally identical to \eqref{eqIso1}, although the two Hamiltonians act on different Hilbert spaces. Nevertheless, most properties of the SU(3) triplet Hamiltonian resemble those of the SU(2) doublet Hamiltonian. 

The band structure corresponding to the Hamiltonian in Eq.~\eqref{eqIso16} consists of a triply degenerate band which in polar coordinates $(k,\vartheta)$ of the wave number $\boldsymbol{k}$ is given by
\begin{equation}
	E(k,\vartheta) = \frac{\hbar^2 k^2}{2M} .
	\label{eqIso17}
\end{equation} 
The system is again isotropic in $\boldsymbol{k}$-space.

The momentum space eigenvectors are identical to that of Eq.~\eqref{eqIso6}:
\begin{equation}
	\mathbf{u}(k,\vartheta) = 1 .
	\label{eqIso18}
\end{equation} Similarly, the group velocity field takes the same form as in Eq.~\eqref{eqIso7b}:
\begin{align}
		v_k(k,\vartheta) &= \frac{\hbar k}{M} , 
		\quad 
		v_\vartheta(k,\vartheta) = 0 . 
		\label{eqIso19b}
\end{align} 
The radial component of the group velocity 
is positive for all $k>0$, i.e., it is an electron-like band that is relevant to calculate the AB scattering states in \eqref{eqGen14}.

The $\mathfrak{su}(3)$ Lie algebra corresponding to the SU(3) group is eight-dimensional, whose standard basis consists of the Gell-Mann matrices:
\begin{equation}
	\begin{aligned}
		\hat{\lambda}_1 &= \begin{pmatrix}
			0 & 1 & 0 \\ 1 & 0 & 0 \\ 0 & 0 & 0
		\end{pmatrix} ,\quad &&\hat{\lambda}_2 = \begin{pmatrix}
			0 & -\iu & 0 \\ \iu & 0 & 0 \\ 0 & 0 & 0
		\end{pmatrix} , \\
		\hat{\lambda}_3 &= \begin{pmatrix}
			1 & 0 & 0 \\ 0 & -1 & 0 \\ 0 & 0 & 0
		\end{pmatrix} ,\quad &&\hat{\lambda}_4 = \begin{pmatrix}
			0 & 0 & 1 \\ 0 & 0 & 0 \\ 1 & 0 & 0
		\end{pmatrix} , \\ 
		\hat{\lambda}_5 &= \begin{pmatrix}
			0 & 0 & -\iu \\ 0 & 0 & 0 \\ \iu & 0 & 0
		\end{pmatrix} ,\quad &&\hat{\lambda}_6 = \begin{pmatrix}
			0 & 0 & 0 \\ 0 & 0 & 1 \\ 0 & 1 & 0
		\end{pmatrix} , \\ 
		\hat{\lambda}_7 &= \begin{pmatrix}
			0 & 0 & 0 \\ 0 & 0 & -\iu \\ 0 & \iu & 0
		\end{pmatrix} ,\quad &&\hat{\lambda}_8 = \frac{1}{\sqrt3}\begin{pmatrix}
			1 & 0 & 0 \\ 0 & 1 & 0 \\ 0 & 0 & -2
		\end{pmatrix} .
	\end{aligned}
	\label{eqIso20}
\end{equation} The gauge potential of this problem is given according to Eq.~\eqref{eqGen1} where the SU(3)-flux $\hat{\Phi}$, or equivalently the dimensionless flux $\hat{\alpha}$ can be parameterized with real parameters $\alpha$ and $\beta$ as follows
\begin{equation}
	\hat{\alpha} = \alpha\hat{\lambda}_3 + \sqrt3\beta\hat{\lambda}_8 = \begin{pmatrix}
		\beta + \alpha & 0 & 0 \\ 0 & \beta - \alpha & 0 \\ 0 & 0 & -2\beta
	\end{pmatrix} .
	\label{eqIso21}
\end{equation} 
A more generic flux could be transformed into the diagonal form of Eq.~\eqref{eqIso21} by a change of basis in the $\mathbb{C}^3$ space, i.e., with a global gauge transformation. The eigenvectors of $\hat{\alpha}$ take the form
\begin{equation}
	\mathbf{w}_1 = \begin{pmatrix}
		1 \\ 0 \\ 0 
	\end{pmatrix} ,\quad \mathbf{w}_2 = \begin{pmatrix}
		0 \\ 1 \\ 0 
	\end{pmatrix} ,\quad \mathbf{w}_3 = \begin{pmatrix}
		0 \\ 0 \\ 1 
	\end{pmatrix} ,
	\label{eqIso22}
\end{equation} with eigenvalues $\beta + \alpha$, $\beta - \alpha$, and $-2\beta$, respectively.

The probability density of the particle position corresponding to an arbitrary state $(\Psi_1,\Psi_2,\Psi_3)\in\mathcal{H}$
can be written as
\begin{equation}
	\varrho = \Psi_1^* \Psi_1 + \Psi_2^* \Psi_2 + \Psi_3^* \Psi_3 .
	\label{eqIso23}
\end{equation} 
Applying the results of the general probability current formula of our previous paper~\cite{Nemeth-Cserti:paper} we find
\begin{subequations}
	\begin{align}
		j_x &= \frac{1}{M} \sum_{n=1}^3 \sum_{l=1}^3 \mathrm{Re}\left[\Psi_n^* \big(\hat{\Pi}_{x,nl} \Psi_l\big)\right] , \label{eqIso24a} \\
		j_y &= \frac{1}{M} \sum_{n=1}^3 \sum_{l=1}^3 \mathrm{Re}\left[\Psi_n^* \big(\hat{\Pi}_{y,nl} \Psi_l\big)\right] . 
		\label{eqIso24b}
	\end{align}
\end{subequations}

We are now ready to perform the contour integral in Eq.~\eqref{eqGen14}. Using the Schäfli--Sommerfeld integral formula of the Bessel functions of the first kind ~\cite{HTF:book}, we obtain 
\begin{subequations}
	\label{eqIso25_3egybe}
\begin{equation}
	\begin{aligned}
		\boldsymbol{\Psi}^{(+)}_{1,\boldsymbol{k}}(r,\varphi) = \sum_{m=-\infty}^\infty &(-\iu)^{|m+\beta+\alpha|} J_{|m+\beta+\alpha|}(k r) \\
		&\times\euler^{\iu m(\varphi - \vartheta + \pi)} \begin{pmatrix} 1 \\ 0 \\ 0 \end{pmatrix} ,
	\end{aligned}
	\label{eqIso25}
\end{equation} for $n=1$, and
\begin{equation}
	\begin{aligned}
		\boldsymbol{\Psi}^{(+)}_{2,\boldsymbol{k}}(r,\varphi) = \sum_{m=-\infty}^\infty &(-\iu)^{|m+\beta-\alpha|} J_{|m+\beta-,\alpha|}(k r) \\ &\times\euler^{\iu m(\varphi - \vartheta + \pi)} \begin{pmatrix} 0 \\ 1 \\ 0 \end{pmatrix} ,
	\end{aligned}
	\label{eqIso26}
\end{equation} for $n=2$, and
\begin{equation}
	\begin{aligned}
		\boldsymbol{\Psi}^{(+)}_{3,\boldsymbol{k}}(r,\varphi) = \sum_{m=-\infty}^\infty &(-\iu)^{|m-2\beta|} J_{|m-2\beta|}(k r) \\
		&\times\euler^{\iu m(\varphi - \vartheta + \pi)} \begin{pmatrix} 0 \\ 0 \\ 1 \end{pmatrix} ,
	\end{aligned}
	\label{eqIso27}
\end{equation}
\end{subequations}
for $n=3$. To the best of our knowledge, these results cannot be found in the literature.

\begin{figure}[!hbt]
	\begin{overpic}[scale=0.5]{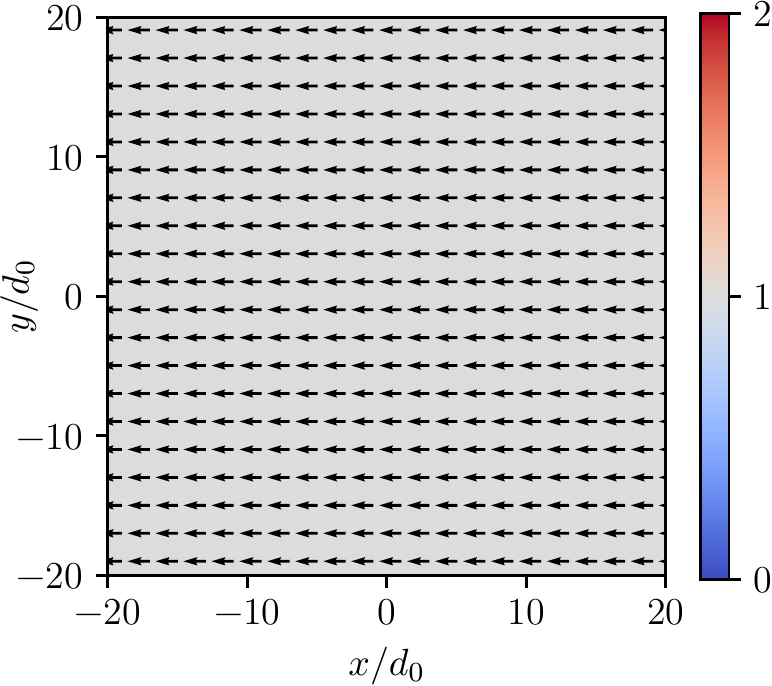}
		\put(-5,80){\scriptsize{(a)}}
	\end{overpic}
	\hspace{10pt}
	\begin{overpic}[scale=0.5]{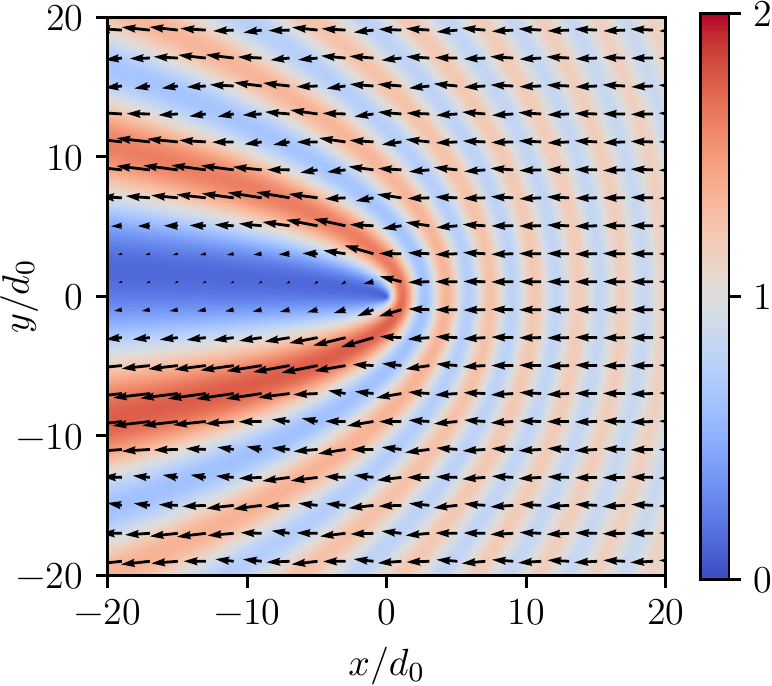}
		\put(-5,80){\scriptsize{(b)}}
	\end{overpic}
	\begin{overpic}[scale=0.5]{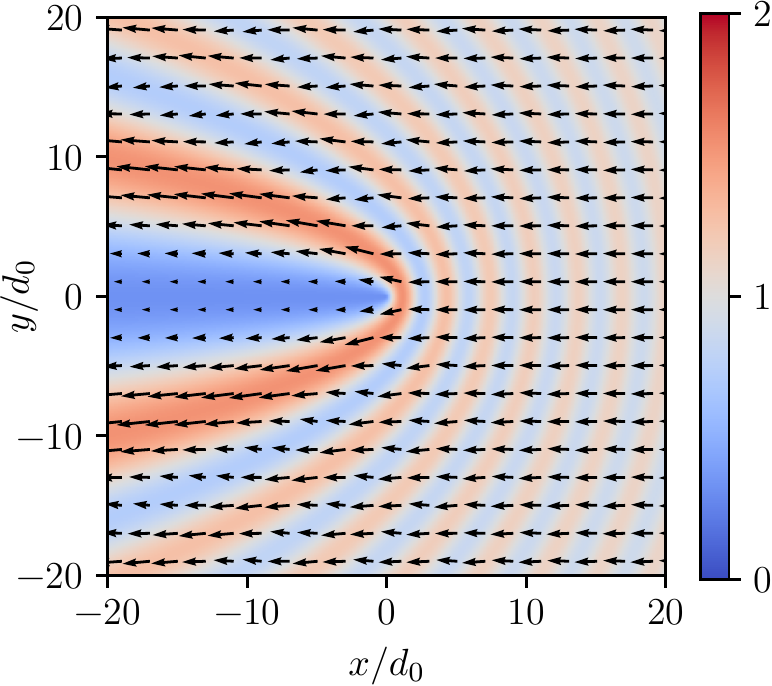}
		\put(-5,80){\scriptsize{(c)}}
	\end{overpic}
	\hspace{10pt}
	\begin{overpic}[scale=0.5]{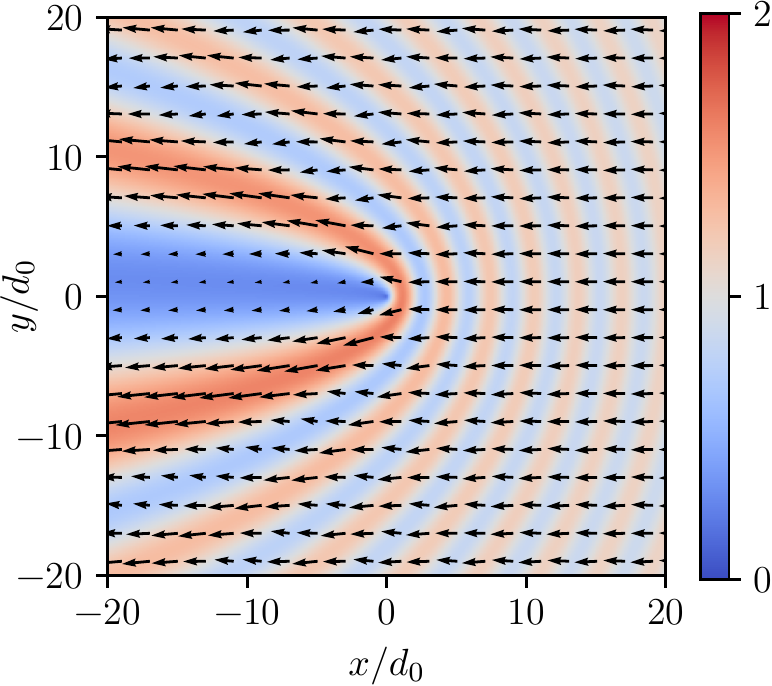}
		\put(-5,80){\scriptsize{(d)}}
	\end{overpic}
    \begin{overpic}[scale=0.5]{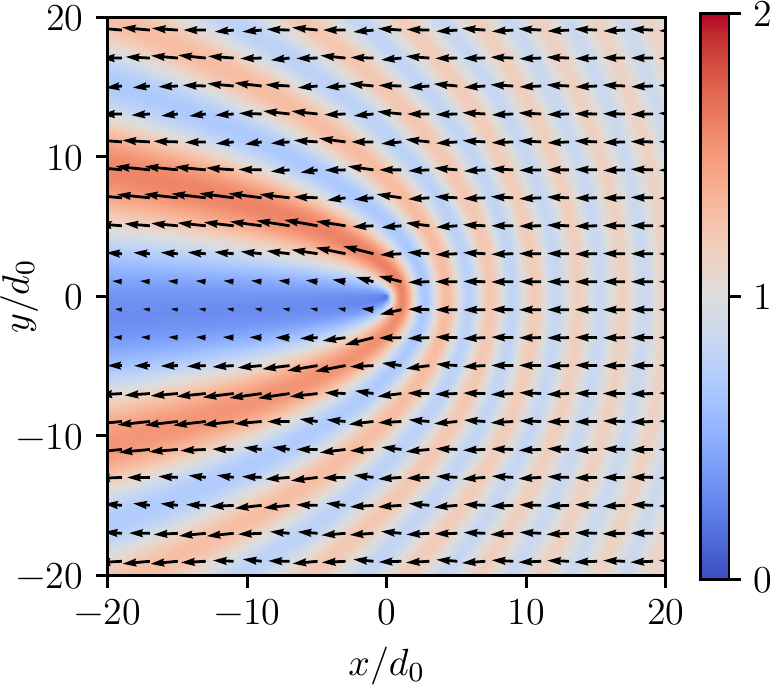}
		\put(-5,80){\scriptsize{(e)}}
	\end{overpic}
	\hspace{10pt}
	\begin{overpic}[scale=0.5]{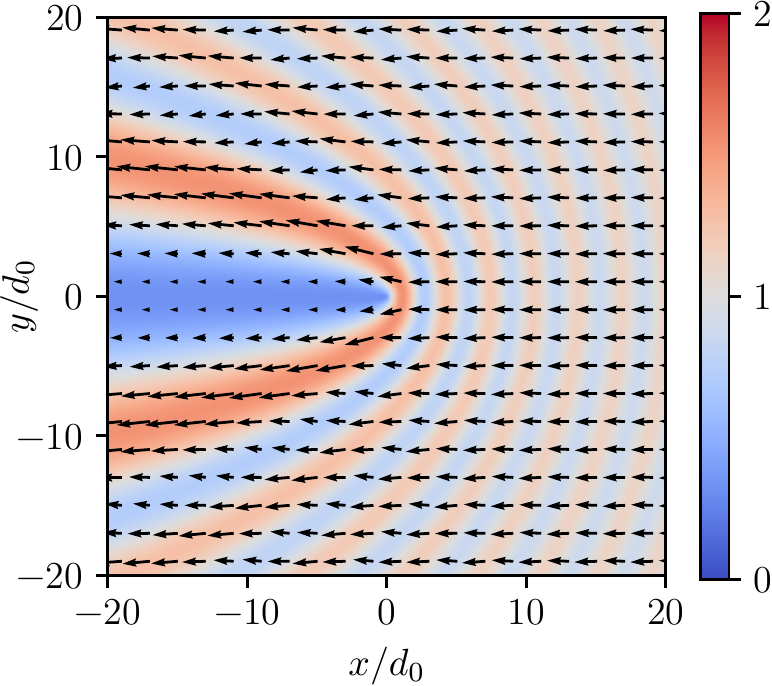}
		\put(-5,80){\scriptsize{(f)}}
	\end{overpic}
	\begin{overpic}[scale=0.5]{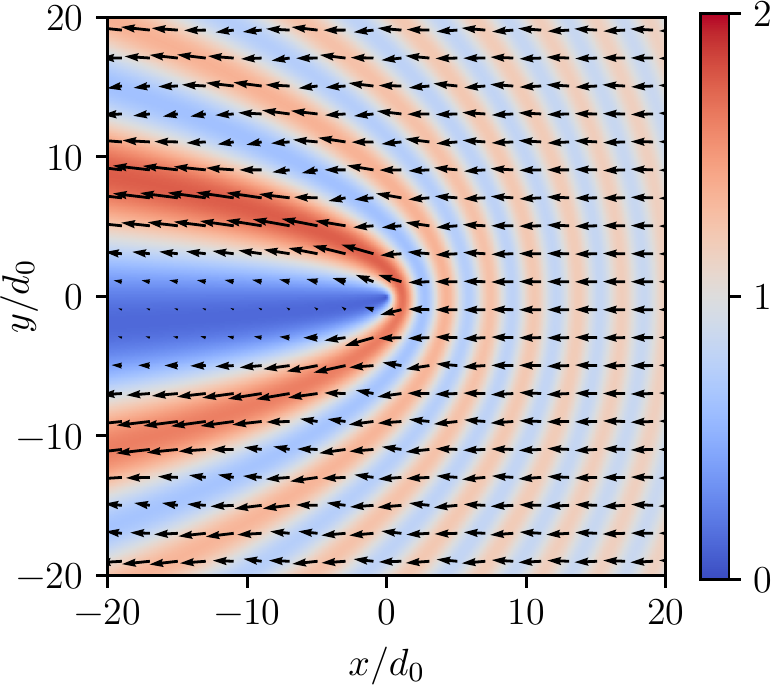}
		\put(-5,80){\scriptsize{(g)}}
	\end{overpic}
	\hspace{10pt}
	\begin{overpic}[scale=0.5]{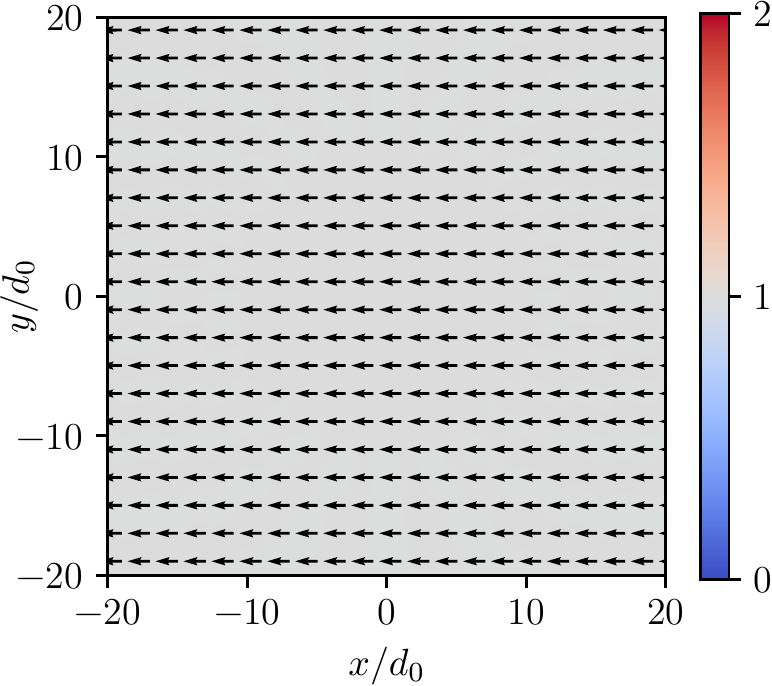}
		\put(-5,80){\scriptsize{(h)}}
	\end{overpic}
	\caption{Scattering states corresponding to the Aharonov--Bohm effect for the SU(3) triplet. The probability density $\varrho$ (represented by the colors) and current density $\boldsymbol{j}$ (represented by the arrows) are computed for $kd_0 = 1$ (where $d_0$ is a natural length unit) and (a) $\alpha = \beta = 0$, (b) $\alpha = 0$, $\beta = 1/3$, (c) $\alpha = 0$, $\beta = 0.5$, (d) $\alpha = 0.2$, $\beta = 1/3$, (e) $\alpha = 0.3$, $\beta = 1/6$, (f) $\alpha = 0.5$, $\beta = 0$, (g) $\alpha = 0.5$, $\beta = 1/6$, (h) $\alpha = 0.5$, $\beta = 0.5$.}
	\label{figColour1}
\end{figure}

We now consider a scattering state built as a linear combination of the wave functions in \eqref{eqIso25_3egybe}:
\begin{equation}
	\boldsymbol{\Psi}^{(+)} = \frac{1}{\sqrt3}\left[\boldsymbol{\Psi}^{(+)}_{1,\boldsymbol{k}} + \boldsymbol{\Psi}^{(+)}_{2,\boldsymbol{k}} + \boldsymbol{\Psi}^{(+)}_{3,\boldsymbol{k}}\right] .
	\label{eqIso28a}
\end{equation} 
To see the features of this scattering state, we calculate the corresponding probability density and current density using Eqs.~\eqref{eqIso23}, \eqref{eqIso24a} and \eqref{eqIso24b}. The results of our numerical calculations are shown in Fig.~\ref{figColour1}.

As expected, they mostly resemble the plots acquired in the Abelian case and for the SU(2) doublet; let us now point out the differences. Depending on the values of $\alpha$ and $\beta$, the reflection symmetry with respect to the $x$ axis is either present or not. Examples of the former can be seen in Figs.~\ref{figColour1}(c) or (f), whereas examples of the latter are in Figs.~\ref{figColour1}(b) or (g).

Another interesting feature is seen in Fig.~\ref{figColour1}(h), namely, for $\alpha = \beta = 0.5$, no scattering is observed, similarly to the case of integer $\alpha$ in the Abelian Aharonov--Bohm effect. This is of course not at all surprising since with these parameters all the three eigenvalues of the  SU(3)-flux $\hat{\alpha}$ in Eq.~\eqref{eqIso21} are integers, leading to the absence of scattering in all components.

\begin{figure}[!hbt]
	\begin{overpic}[scale=0.5]{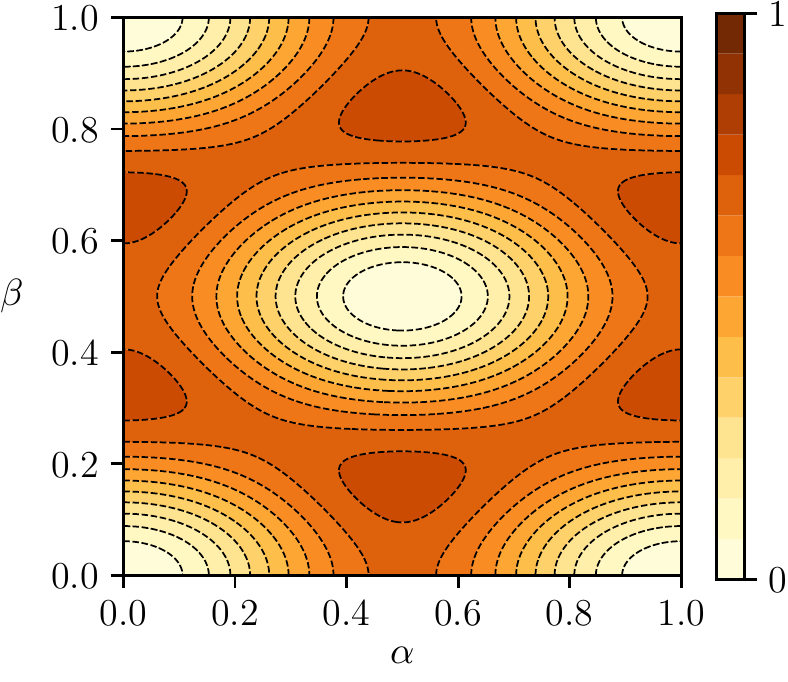}
		\put(-5,80){\scriptsize{(a)}}
	\end{overpic}
	\hspace{5pt}
	\begin{overpic}[scale=0.5]{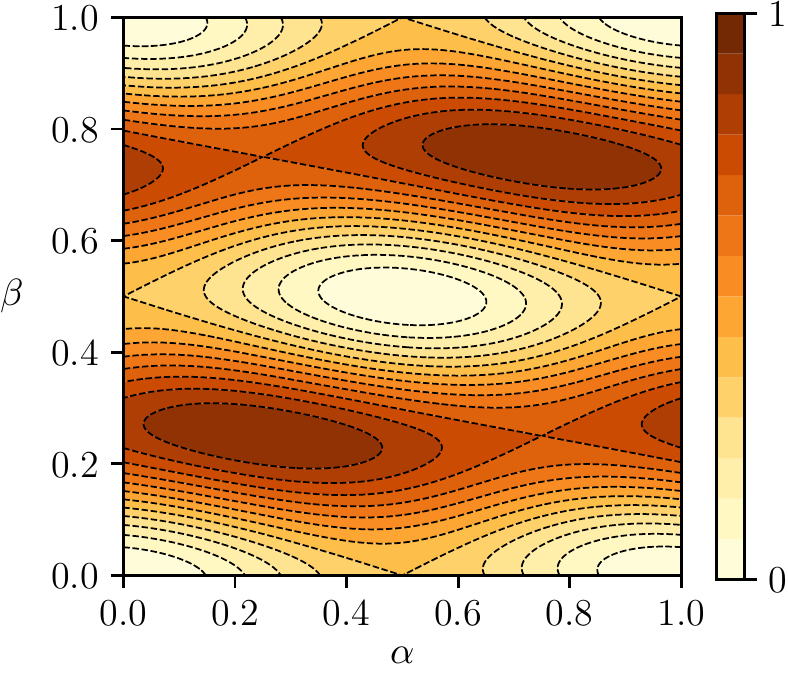}
		\put(-5,80){\scriptsize{(b)}}
	\end{overpic}
	\begin{overpic}[scale=0.5]{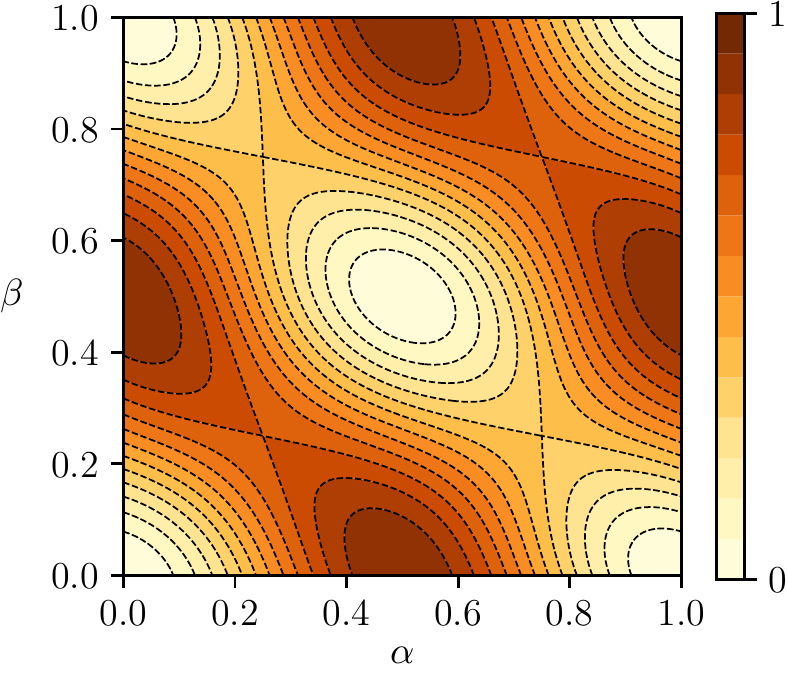}
		\put(-5,80){\scriptsize{(c)}}
	\end{overpic}
	\hspace{5pt}
	\begin{overpic}[scale=0.5]{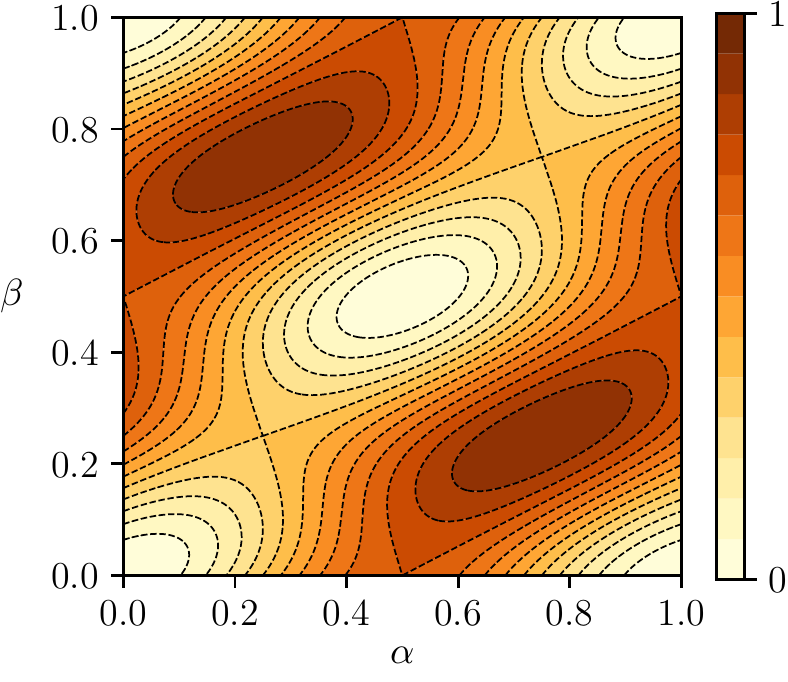}
		\put(-5,80){\scriptsize{(d)}}
	\end{overpic}
	\caption{Dimensionless cross section factor for the SU(3) triplet as a function of $\alpha$ and $\beta$ for (a) $c_1 : c_2 : c_3 = 1 : 1 : 1$, (b) $c_1 : c_2 : c_3 = 1 : 2 : 3$, (c) $c_1 : c_2 : c_3 = 2 : 3 : 1$, (d) $c_1 : c_2 : c_3 = 3 : 1 : 2$. The relation $|c_1|^2 + |c_2|^2 + |c_3|^2 = 1$ is always satisfied.}
	\label{figColour2}
\end{figure}

Finally, we calculate the cross section factor $\Sigma$ defined in Eq.~\eqref{eqGen18_b} for a general incoming plane wave according to Eq.~\eqref{eqGen15} using the states in Eqs.~\eqref{eqIso25_3egybe}, and we obtain
\begin{equation}
	\begin{aligned}
		\Sigma &= |c_1|^2\sin^2\left[(\beta + \alpha)\pi\right] \\ &+ |c_2|^2\sin^2\left[(\beta - \alpha)\pi\right] \\ &+ |c_3|^2\sin^2\left[2\beta\pi\right] ,
	\end{aligned}
	\label{eqColour16}
\end{equation} 
which now depends on the linear combination coefficients $c_n$. 
A few examples of $\Sigma$ for incident waves with different $c_n$ values are shown in Fig.~\ref{figColour2}. 

We now discuss the details of these results. The most symmetric case is shown in Fig.~\ref{figColour2}(a) where the three polarizations are present with equal weight. One can see that the hexagonal structure in the magnitude of $\Sigma$ is distorted as the coefficients $c_n$ are changed as shown in Figs.~\ref{figColour2}(b), (c), and (d). 
Irrespective of weights $c_n$, the differential cross section shows periodicity with period $1$ in both $\alpha$ and $\beta$, and it vanishes at $\alpha,\beta\in\mathbb{R}$, similarly to the Abelian case. 
However, there is also a nontrivial zero value of the cross section factor $\Sigma$ at $\alpha = \beta = 0.5$, in agreement with Fig.~\ref{figColour1}(h). 
Finally, we should mention that the maximal value of $\Sigma$ does generally not reach unity, which is best seen in Fig.~\ref{figColour2}(a), but can be observed in other cases as well.

\subsection{U(2) doublet}
\label{sec33}

The third example we consider here is the \emph{U(2) doublet}. From our point of view, this is equivalent to its locally isomorphic counterpart, the SU(2)$\times$U(1) doublet, appearing in a model of $K^+$ and $K^0$ kaons where they are regarded as different isospin--hypercharge eigenstates of the same particle ~\cite{Cheng:1984vwu,Nachtmann:1990ta,Horvath:2019atn}. For the sake of simplicity and brevity, we use the ultrarelativistic limit, i.e., the masses of kaons are neglected in our model.  
However, we should note that the generic relativistic regime can also be treated in our theory. An analogous system can be engineered using again ultracold atoms~\cite{Osterloh:paper,Jacob:2007jb} trapped in an optical honeycomb lattice  mimicking monolayer graphene~\cite{Bermudez:paper}. In the low-energy regime of this system, the dispersion relation resembles that of a massless relativistic particle. To the best of our knowledge, the calculation of the AB scattering states of this system were not performed before. In what follows, we apply our  methods developed in Sec.~\ref{sec2} to study the nature of the scattering wave functions and the differential cross section.

The Hilbert space corresponding to the U(2) doublet is $\mathcal{H} = L^2(\mathbb{R}^2,\mathbb{C})\otimes\mathbb{C}^4$, that is, there is a two-dimensional internal degree of freedom ($D=2$), and a two-dimensional gauge degree of freedom ($N=2$) corresponding to polarization. The Hamiltonian operator $\hat{H}: \mathcal{D}_H \to \mathcal{H}$ is given as a linear polynomial of the momentum operators:
\begin{equation}
	\hat{H} = v \big(\hat{p}_x\otimes\hat{I}_2\otimes\hat{\sigma}_x + \hat{p}_y\otimes\hat{I}_2\otimes\hat{\sigma}_y\big) ,
	\label{eqHyper1}
\end{equation} where $v$ is an effective velocity parameter, and 
\begin{equation}
	\hat{\sigma}_x = \begin{pmatrix}
		0 & 1 \\ 1 & 0
	\end{pmatrix} ,\quad \hat{\sigma}_y = \begin{pmatrix}
		0 & -\mathrm{i} \\ \mathrm{i} & 0
	\end{pmatrix} ,\quad \hat{\sigma}_z = \begin{pmatrix}
		1 & 0 \\ 0 & -1
	\end{pmatrix}
	\label{eqHyper2}
\end{equation} are the Pauli matrices. 

The band structure corresponding to the Hamiltonian in Eq.~\eqref{eqHyper1} consists of two doubly degenerate bands with band index $s\in\{-1,1\}$ which in $\boldsymbol{k}$-space polar coordinates is given by 
\begin{equation}
	E_s(k,\vartheta) = s v \hbar k .
	\label{eqHyper6}
\end{equation} 
As before, the dispersion relation is isotropic.

The momentum space eigenvectors need to be chosen such that one of their components is independent of $\vartheta$.
This is related to the regularity of the wave function at the origin as it was discussed in detail in our previous work~\cite{Nemeth-Cserti:paper}. 
Thus, for simplicity, we choose the following eigenvector:
\begin{equation}
	\mathbf{u}_s(k,\vartheta) = \frac{1}{\sqrt2} \begin{pmatrix}
		1 \\ s \euler^{\iu\vartheta}
	\end{pmatrix} .
	\label{eqHyper8}
\end{equation} 

Then, from Eqs.~\eqref{group_velocity:eq} and \eqref{eqHyper6} the group velocity is given by
\begin{align}
		v_{s,k}(k,\vartheta) &= sv , 
		\quad v_{s,\vartheta}(k,\vartheta) = 0 . 
		\label{eqHyper9b}
\end{align} 
Now, it is clear that the radial group velocity for all $k$ is either positive or negative depending on the band index $s$. This implies that the band with $s=1$ is electron-like whereas the band with $s=-1$ is hole-like. 

The $\mathfrak{u}(2)$ Lie algebra corresponding to the $\mathrm{U}(2)$ group is four-dimensional, whose standard basis consists of the half-Pauli and half-identity matrices:
\begin{equation}
	\begin{aligned}
		\hat{\tau}_1 &= \frac12 \begin{pmatrix}
			0 & 1 \\ 1 & 0
		\end{pmatrix} ,\quad &&\hat{\tau}_2 = \frac12 \begin{pmatrix}
			0 & -\iu \\ \iu & 0
		\end{pmatrix} , \\ 
		\hat{\tau}_3 &= \frac12 \begin{pmatrix}
			1 & 0 \\ 0 & -1
		\end{pmatrix},\quad &&\hat{\tau}_4 = \frac12 \begin{pmatrix}
			1 & 0 \\ 0 & 1
		\end{pmatrix} .
	\end{aligned}
	\label{eqHyper10}
\end{equation} The gauge potential of this problem is given according to Eq.~\eqref{eqGen1} where the $\mathrm{U}(2)$-flux $\hat{\Phi}$, or equivalently the dimensionless flux $\hat{\alpha}$ can be parameterized with real parameters $\alpha$ and $\beta$ as follows
\begin{equation}
	\hat{\alpha} = 2\alpha\hat{\tau}_3 + 2\beta\hat{\tau}_4 = \begin{pmatrix}
		\beta + \alpha & 0 \\ 0 & \beta - \alpha
	\end{pmatrix} ,
	\label{eqHyper11}
\end{equation}
and the eigenvectors of $\hat{\alpha}$ are
\begin{equation}
	\mathbf{w}_1 = \begin{pmatrix}
		1 \\ 0
	\end{pmatrix} ,\quad \mathbf{w}_2 = \begin{pmatrix}
		0 \\ 1
	\end{pmatrix} ,
	\label{eqHyper12}
\end{equation} with eigenvalues $\beta + \alpha$ and $\beta - \alpha$, respectively.
 A more generic flux could be transformed into the diagonal form of Eq.~\eqref{eqHyper11} by a change of basis in the $\mathbb{C}^2$ space, i.e., a global gauge transformation.

Then, the probability density of the particle position corresponding to an arbitrary state $\boldsymbol{\Psi} = \left(\Psi_1,\Psi_2,\Psi_3,\Psi_4\right) \in \mathcal{H}$ can be written as
\begin{equation}
	\varrho = \Psi_1^* \Psi_1 + \Psi_2^* \Psi_2 + \Psi_3^* \Psi_3 + \Psi_4^* \Psi_4 .
	\label{eqHyper13}
\end{equation} 
Now, using the results of the general probability current formula in our previous paper~\cite{Nemeth-Cserti:paper}, we find
\begin{subequations}
	\label{eqHyper14}
	\begin{align}
		j_x &= 2v~\mathrm{Re}\left(\Psi_1^*\Psi_2 + \Psi_3^*\Psi_4\right) , \label{eqHyper14a} \\
		j_y &= 2v~\mathrm{Im}\left(\Psi_1^*\Psi_2 + \Psi_3^*\Psi_4\right) . 
		\label{eqHyper14b}
	\end{align}
\end{subequations}

Performing the contour integral in Eq.~\eqref{eqGen14} using again the Schäfli--Sommerfeld integral formula of the Bessel functions of the first kind ~\cite{HTF:book},  we obtain the following scattering wave functions
\begin{subequations}
	\label{eqHyper15-18_egybe}
\begin{equation}
	\begin{aligned}
		&\boldsymbol{\Psi}^{(+)}_{1,1,\boldsymbol{k}}(r,\varphi) = \sum_{m=-\infty}^\infty \hspace{-4pt}\frac{(-\iu)^{|m+\beta+\alpha|} \euler^{\iu m (\varphi - \vartheta + \pi)}}{\sqrt2}  \\[2ex] 
		&\times\begin{pmatrix} J_{|m+\beta+\alpha|}(k r) \\ \iu \epsilon(m + \beta + \alpha) J_{|m+\beta+\alpha| + \epsilon(m+\beta+\alpha)}(k r) \euler^{\iu\varphi} \\ 0 \\ 0 \end{pmatrix} ,
	\end{aligned}
	\label{eqHyper15}
\end{equation}  for $s=1$, $n=1$, 
\begin{equation}
	\begin{aligned}
		&\boldsymbol{\Psi}^{(+)}_{-1,1,\boldsymbol{k}}(r,\varphi) = \sum_{m=-\infty}^\infty \hspace{-4pt}\frac{\iu^{|m+\beta+\alpha|} \euler^{\iu m (\varphi - \vartheta + \pi)}}{\sqrt2} \\ 
		&\hspace{-6pt}\times\begin{pmatrix} J_{|m+\beta+\alpha|}(k r) \\ -\iu \epsilon(m + \beta + \alpha) J_{|m+\beta+\alpha| + \epsilon(m+\beta+\alpha)}(k r) \euler^{\iu\varphi} \\ 0 \\ 0 \end{pmatrix} ,
	\end{aligned}
	\label{eqHyper16}
\end{equation}
for $s=-1$, $n=1$,
\begin{equation}
	\begin{aligned}
		&\boldsymbol{\Psi}^{(+)}_{1,2,\boldsymbol{k}}(r,\varphi) = \sum_{m=-\infty}^\infty \hspace{-4pt}\frac{(-\iu)^{|m+\beta-\alpha|} \euler^{\iu m (\varphi - \vartheta + \pi)}}{\sqrt2} \\
		&\times\begin{pmatrix} 0 \\ 0 \\ J_{|m+\beta-\alpha|}(k r) \\ \iu \epsilon(m + \beta - \alpha) J_{|m+\beta-\alpha| + \epsilon(m+\beta-\alpha)}(k r) \euler^{\iu\varphi} \end{pmatrix} ,
	\end{aligned}
	\label{eqHyper17}
\end{equation}
for $s=1$, $n=2$, and 
\begin{equation}
	\begin{aligned}
		&\boldsymbol{\Psi}^{(+)}_{-1,2,\boldsymbol{k}}(r,\varphi) = \sum_{m=-\infty}^\infty \hspace{-4pt}\frac{\iu^{|m+\beta-\alpha|} \euler^{\iu m (\varphi - \vartheta + \pi)}}{\sqrt2} \\ 
		&\hspace{-6pt}\times\begin{pmatrix} 0 \\ 0 \\ J_{|m+\beta-\alpha|}(k r) \\ -\iu \epsilon(m + \beta - \alpha) J_{|m+\beta-\alpha| + \epsilon(m+\beta-\alpha)}(k r) \euler^{\iu\varphi} \end{pmatrix} ,
	\end{aligned}
	\label{eqHyper18}
\end{equation}
\end{subequations} 
for $s=-1$, $n=2$.
This result is another non-trivial application of our general method, and to the best of our knowledge, it is not yet published in the literature.

\begin{figure}[!hbt]
	\begin{overpic}[scale=0.5]{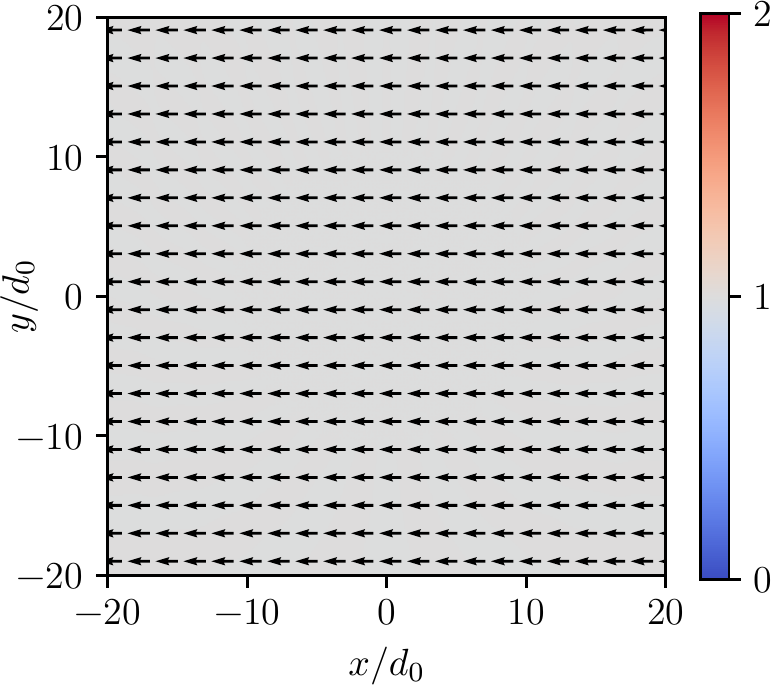}
		\put(-5,80){\scriptsize{(a)}}
	\end{overpic}
	\hspace{10pt}
	\begin{overpic}[scale=0.5]{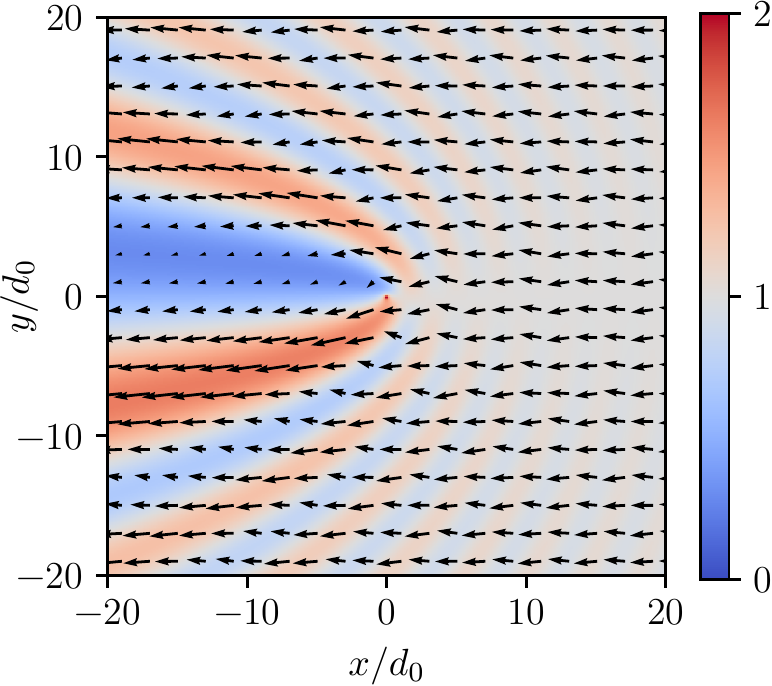}
		\put(-5,80){\scriptsize{(b)}}
	\end{overpic}
	\begin{overpic}[scale=0.5]{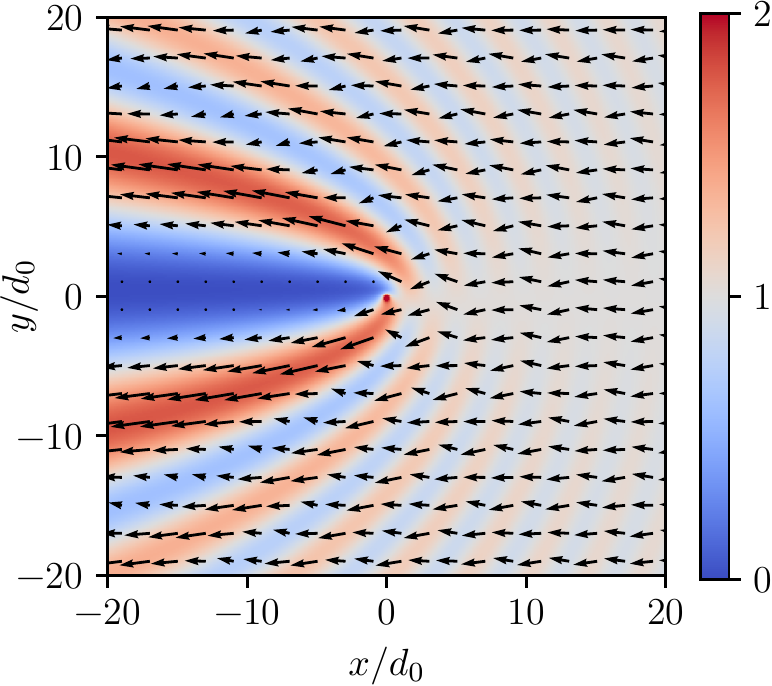}
		\put(-5,80){\scriptsize{(c)}}
	\end{overpic}
	\hspace{10pt}
	\begin{overpic}[scale=0.5]{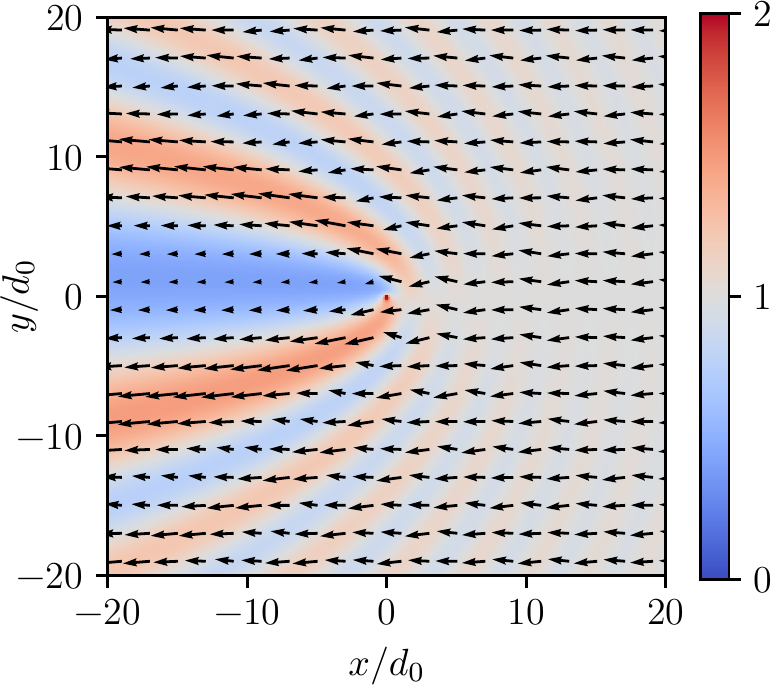}
		\put(-5,80){\scriptsize{(d)}}
	\end{overpic}
	\begin{overpic}[scale=0.5]{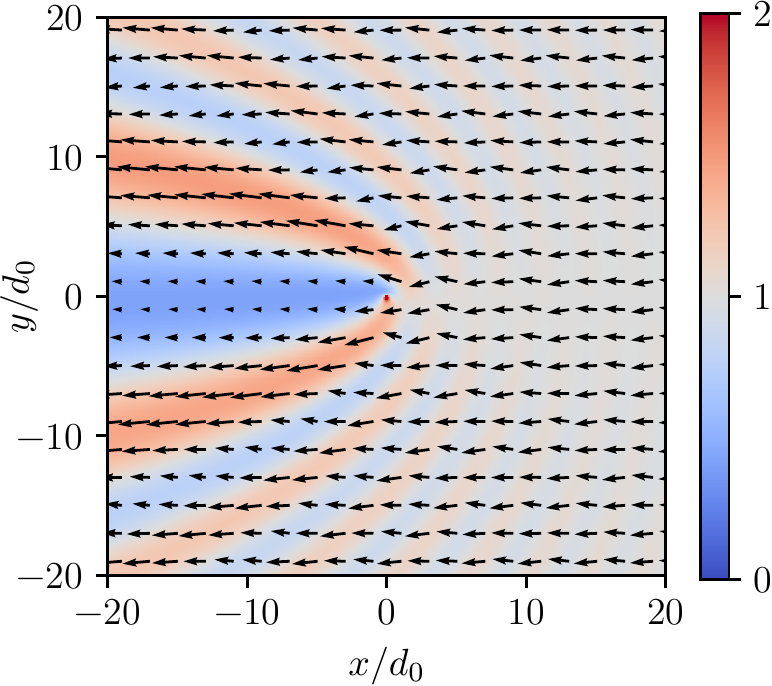}
		\put(-5,80){\scriptsize{(e)}}
	\end{overpic}
	\hspace{10pt}
	\begin{overpic}[scale=0.5]{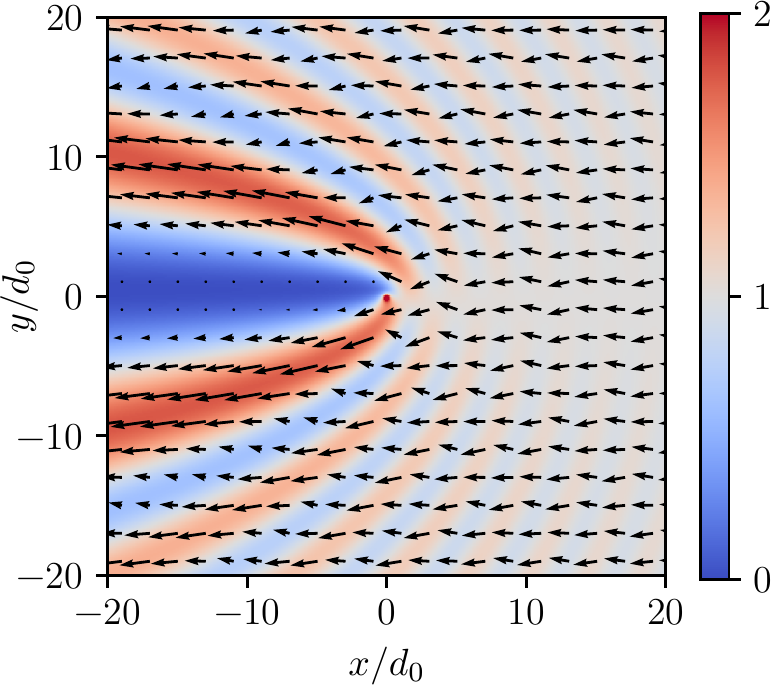}
		\put(-5,80){\scriptsize{(f)}}
	\end{overpic}
	\begin{overpic}[scale=0.5]{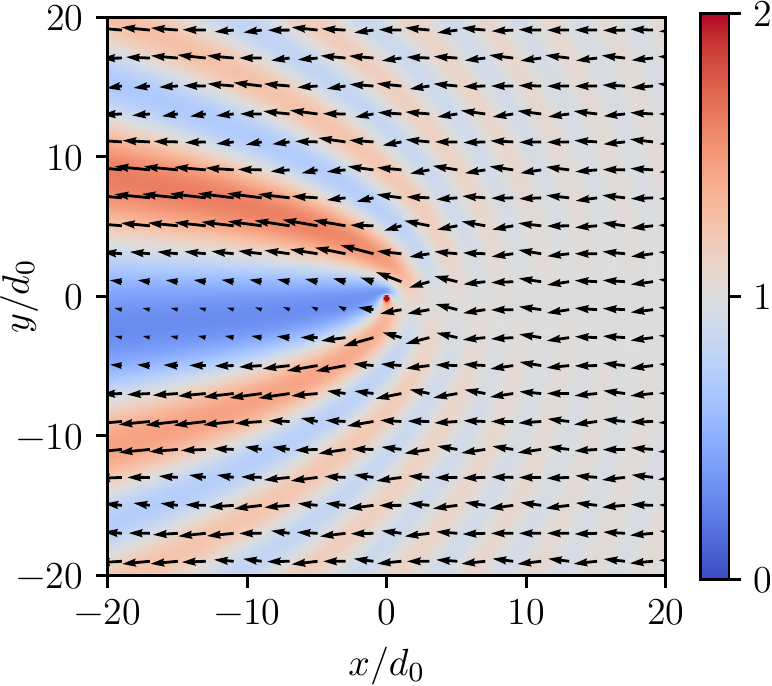}
		\put(-5,80){\scriptsize{(g)}}
	\end{overpic}
	\hspace{10pt}
	\begin{overpic}[scale=0.5]{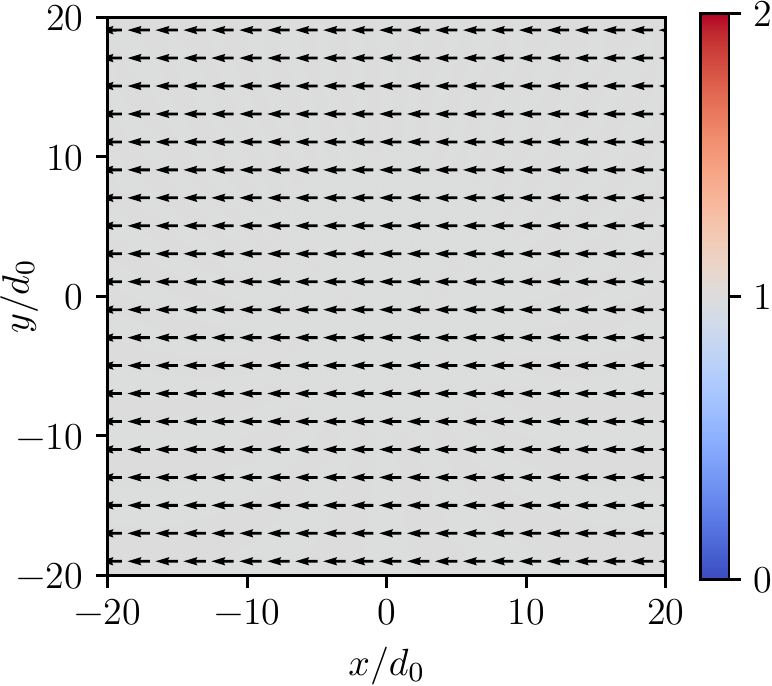}
		\put(-5,80){\scriptsize{(h)}}
	\end{overpic}
	\caption{Scattering states corresponding to the Aharonov--Bohm effect for the U(2) doublet. The probability density $\varrho$ (represented by the colors) and current density $\boldsymbol{j}$ (represented by the arrows) are computed for $kd_0 = 1$ (where $d_0$ is a natural length unit) and (a) $\alpha = \beta = 0$, (b) $\alpha = 0$, $\beta = 0.25$, (c) $\alpha = 0$, $\beta = 0.5$, (d) $\alpha = 0.2$, $\beta = 0.3$, (e) $\alpha = 0.3$, $\beta = 0.2$, (f) $\alpha = 0.5$, $\beta = 0$, (g) $\alpha = 0.5$, $\beta = 0.25$, (h) $\alpha = 0.5$, $\beta = 0.5$.}
	\label{figHyper2}
\end{figure}

Now, we consider a special wave function built as a linear combination of Eqs.~\eqref{eqHyper15} and \eqref{eqHyper17}:
\begin{equation}
	\boldsymbol{\Psi}^{(+)} = \frac{1}{\sqrt2}\left[\boldsymbol{\Psi}^{(+)}_{1,1,\boldsymbol{k}} + \boldsymbol{\Psi}^{(+)}_{1,2,\boldsymbol{k}}\right] .
	\label{eqHyper19}
\end{equation} 
Now, using again Eqs.~\eqref{eqHyper13} and \eqref{eqHyper14}, we calculate the probability density and current density, respectively, and the results are shown in Fig.~\ref{figHyper2}.
As we expected, these are similar to the wave functions of the Abelian case and the SU(2) doublet. However, a few differences should be pointed out.
First, in Fig.~\ref{figHyper2}(h)  for parameters $\alpha = \beta = 0.5$, no scattering is observed, similarly to the SU(3) triplet discussed in Sec.~\ref{sec32}.
This is clear because for these parameters the eigenvalues of the dimensionless $\mathrm{U}(2)$-flux $\hat{\alpha}$ in Eq.~\eqref{eqHyper11} are integers resulting in the absence of scattering in all components.

\begin{figure}[!hbt]
	\begin{overpic}[scale=0.5]{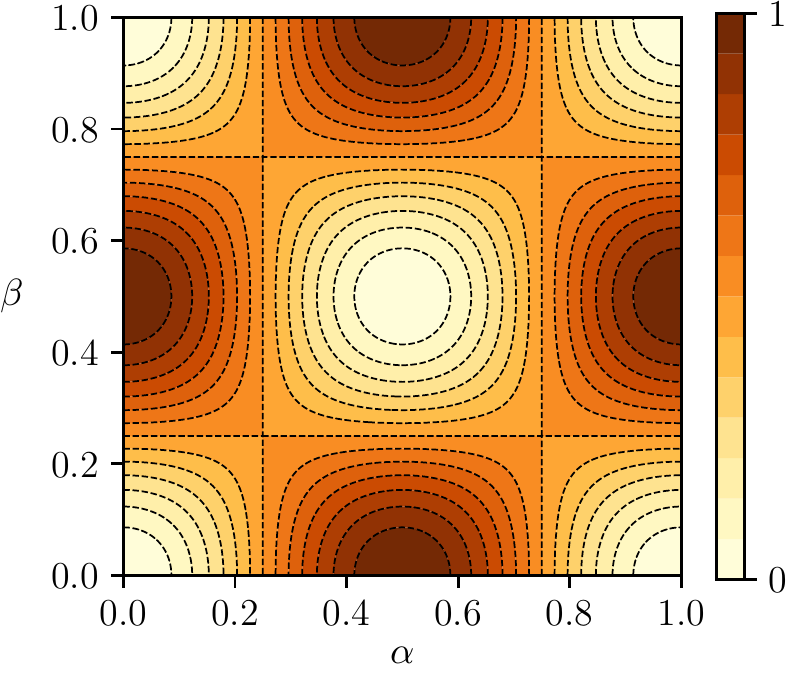}
		\put(-5,80){\scriptsize{(a)}}
	\end{overpic}
	\hspace{5pt}
	\begin{overpic}[scale=0.5]{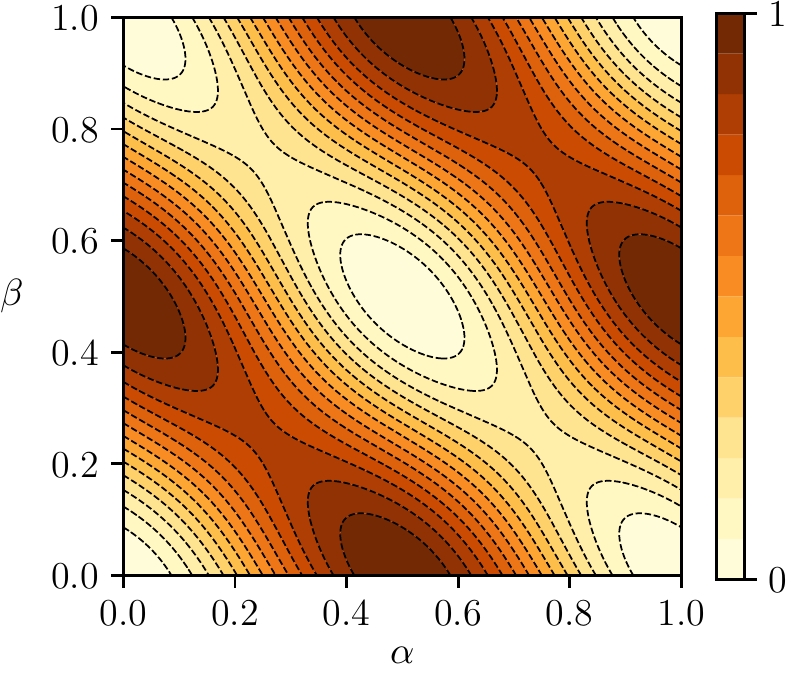}
		\put(-5,80){\scriptsize{(b)}}
	\end{overpic}
	\begin{overpic}[scale=0.5]{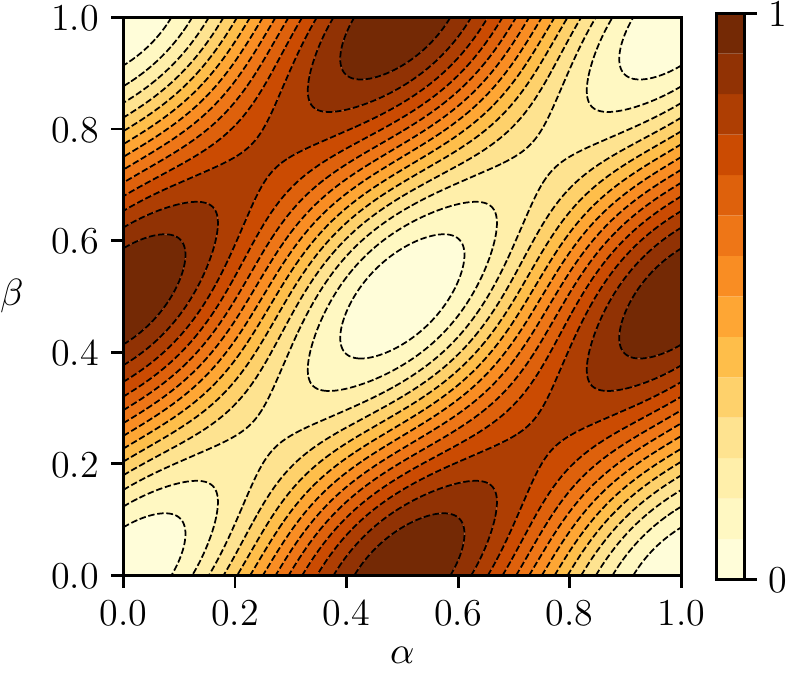}
		\put(-5,80){\scriptsize{(c)}}
	\end{overpic}
	\caption{Dimensionless cross section factor for the U(2) doublet as a function of $\alpha$ and $\beta$ for (a) $c_1 : c_2 = 1 : 1$, (b) $c_1 : c_2 = 1 : 2$, (c) $c_1 : c_2 = 2 : 1$. The relation $|c_1|^2 + |c_2|^2 = 1$ is always satisfied.}
	\label{figHyper3}
\end{figure}

Finally, we calculate the cross section factor $\Sigma$ defined in Eq.~\eqref{eqGen18} 
for  the U(2) doublet and obtain
\begin{equation}
	\Sigma = |c_1|^2\sin^2\left[(\beta + \alpha)\pi\right] + |c_2|^2\sin^2\left[(\beta - \alpha)\pi\right] ,
    \label{eqHyper21}
\end{equation} which is again dependent on the linear combination coefficients $c_n$. A few examples regarding different incident waves are shown in Fig.~\ref{figHyper3}. 

The following general features of $\Sigma$ can be observed as the coefficients $c_n$ of the incoming waves are changed. 
The tetragonal structure in the magnitude of $\Sigma$ shown in  Fig.~\ref{figHyper3}(a) will be distorted as the ratio of $c_1/c_2$ is changed as it is clearly seen in Figs.~\ref{figHyper3}(b) and (c). 
Again, irrespective of $c_1$ and $c_2$, the differential cross section shows a periodicity with period $1$ in both $\alpha$ and $\beta$, and it vanishes at $\alpha,\beta\in\mathbb{R}$ and $\alpha = \beta = 0.5$, similarly to the SU(3) triplet.
Moreover, the maximal value of the cross section factor $\Sigma$ is always unity if $\alpha + \beta$ and $\alpha - \beta$ are half-integers. Actually, this feature is rather different from the case of the SU(3) triplet. 
Finally, an interesting feature can be observed by comparing Figs.~\ref{figHyper3}(b) and (c): those are reflections of each other as a result of the symmetric partition of weights, that is, $c_1/c_2 = 2$ or $c_2/c_1 = 2$.

\section{Conclusions}

To conclude, for the non-Abelian Aharonov--Bohm effect in isotropic multiband systems we have two central results: first, the contour integral representation of the scattering states; and second, the differential scattering cross section; given in Eqs.~\eqref{eqGen14} and \eqref{eqGen18}, respectively.
The form of the scattering states is a natural extension of that established for the Abelian AB problem in our recent work~\cite{Nemeth-Cserti:paper}. 
As far as we know, our theory of the non-Abelian AB problem for isotropic multiband systems is a new result in the literature. 

Similarly to the Abelian case, our proposed wave functions satisfy the necessary boundary and regularity conditions.  
Moreover, using the asymptotic form of the non-Abelian scattering states given by~\eqref{eqGen14}, we obtained the differential scattering cross section which has the following main features.
First, the angular dependence of the cross section is the \emph{same} as that obtained originally by Aharonov and Bohm in their seminal paper~\cite{Aharonov-Bohm:cikk}. 
Second, in contrast to the Abelian case, we found that the cross section now depends on the polarization of the incoming plane wave taking into account the so-called cross section factor $\Sigma$ given by Eq.~\eqref{eqGen18_b}. 

To show how versatile our theory is, in Sec.~\ref{sec3} we carried out the complex contour integrals in Eq.~\eqref{eqGen14} for three non-trivial isotropic multiband systems relevant to condensed matter and particle physics. For each case, we obtained the explicit form of the scattering states, and, to have a deeper insight into their nature, we plotted their probability and current distributions for different incoming waves. We pointed out the similarities and differences between these wave functions and those of the Abelian case.  

Our first example, the SU(2) doublet serves as a test problem since it has already been studied by Wu and Yang~\cite{Wu:NA}, and by Horváthy~\cite{Horvathy:NA}. Indeed, our general approach gives the \textit{same} scattering states as those derived by Horváthy. 
Then, in Secs.~\ref{sec32} and~\ref{sec33} we calculated the wave functions for two further systems: the SU(3) triplet and the U(2) doublet, respectively. 
Furthermore, we calculated the cross section factor $\Sigma$ for different incoming polarizations in these systems. The results show a periodic structure as a function of the flux parameters.    
Our general theory can easily be applied to other non-Abelian isotropic multiband systems. 
We believe that our work broadens the scope of the exactly solvable Aharonov-Bohm scattering problems.

\section*{Acknowledgments}

We thank M.~Berry, P. A. Horváthy, C.~Lambert, B.~D\'ora, and Gy.~D\'avid  for valuable discussions.  
This research was supported by the Ministry of Culture and Innovation and the National Research,
Development and Innovation Office within the Quantum Information National Laboratory of Hungary
(Grant No. 2022-2.1.1-NL-2022-00004), by the ÚNKP-22-3 New National Excellence Program of the Ministry for Culture and Innovation from the Source of the National Research, Development and Innovation Fund, and by the Innovation Office (NKFIH) through Grant Nos. K134437.


\begin{thebibliography}{35}%
\makeatletter
\providecommand \@ifxundefined [1]{%
 \@ifx{#1\undefined}
}%
\providecommand \@ifnum [1]{%
 \ifnum #1\expandafter \@firstoftwo
 \else \expandafter \@secondoftwo
 \fi
}%
\providecommand \@ifx [1]{%
 \ifx #1\expandafter \@firstoftwo
 \else \expandafter \@secondoftwo
 \fi
}%
\providecommand \natexlab [1]{#1}%
\providecommand \enquote  [1]{``#1''}%
\providecommand \bibnamefont  [1]{#1}%
\providecommand \bibfnamefont [1]{#1}%
\providecommand \citenamefont [1]{#1}%
\providecommand \href@noop [0]{\@secondoftwo}%
\providecommand \href [0]{\begingroup \@sanitize@url \@href}%
\providecommand \@href[1]{\@@startlink{#1}\@@href}%
\providecommand \@@href[1]{\endgroup#1\@@endlink}%
\providecommand \@sanitize@url [0]{\catcode `\\12\catcode `\$12\catcode
  `\&12\catcode `\#12\catcode `\^12\catcode `\_12\catcode `\%12\relax}%
\providecommand \@@startlink[1]{}%
\providecommand \@@endlink[0]{}%
\providecommand \url  [0]{\begingroup\@sanitize@url \@url }%
\providecommand \@url [1]{\endgroup\@href {#1}{\urlprefix }}%
\providecommand \urlprefix  [0]{URL }%
\providecommand \Eprint [0]{\href }%
\providecommand \doibase [0]{https://doi.org/}%
\providecommand \selectlanguage [0]{\@gobble}%
\providecommand \bibinfo  [0]{\@secondoftwo}%
\providecommand \bibfield  [0]{\@secondoftwo}%
\providecommand \translation [1]{[#1]}%
\providecommand \BibitemOpen [0]{}%
\providecommand \bibitemStop [0]{}%
\providecommand \bibitemNoStop [0]{.\EOS\space}%
\providecommand \EOS [0]{\spacefactor3000\relax}%
\providecommand \BibitemShut  [1]{\csname bibitem#1\endcsname}%
\let\auto@bib@innerbib\@empty
\bibitem [{\citenamefont {Aharonov}\ and\ \citenamefont
  {Bohm}(1959)}]{Aharonov-Bohm:cikk}%
  \BibitemOpen
  \bibfield  {author} {\bibinfo {author} {\bibfnamefont {Y.}~\bibnamefont
  {Aharonov}}\ and\ \bibinfo {author} {\bibfnamefont {D.}~\bibnamefont
  {Bohm}},\ }\bibfield  {title} {\bibinfo {title} {Significance of
  electromagnetic potentials in the quantum theory},\ }\href
  {https://doi.org/10.1103/PhysRev.115.485} {\bibfield  {journal} {\bibinfo
  {journal} {Phys. Rev.}\ }\textbf {\bibinfo {volume} {115}},\ \bibinfo {pages}
  {485} (\bibinfo {year} {1959})}\BibitemShut {NoStop}%
\bibitem [{\citenamefont {Chambers}(1960)}]{PhysRevLett.5.3}%
  \BibitemOpen
  \bibfield  {author} {\bibinfo {author} {\bibfnamefont {R.~G.}\ \bibnamefont
  {Chambers}},\ }\bibfield  {title} {\bibinfo {title} {Shift of an electron
  interference pattern by enclosed magnetic flux},\ }\href
  {https://doi.org/10.1103/PhysRevLett.5.3} {\bibfield  {journal} {\bibinfo
  {journal} {Phys. Rev. Lett.}\ }\textbf {\bibinfo {volume} {5}},\ \bibinfo
  {pages} {3} (\bibinfo {year} {1960})}\BibitemShut {NoStop}%
\bibitem [{\citenamefont {Berry}(1980)}]{Berry_1980}%
  \BibitemOpen
  \bibfield  {author} {\bibinfo {author} {\bibfnamefont {M.~V.}\ \bibnamefont
  {Berry}},\ }\bibfield  {title} {\bibinfo {title} {Exact {Aharonov-Bohm}
  wavefunction obtained by applying {D}irac{'}s magnetic phase factor},\ }\href
  {https://doi.org/10.1088/0143-0807/1/4/011} {\bibfield  {journal} {\bibinfo
  {journal} {European Journal of Physics}\ }\textbf {\bibinfo {volume} {1}},\
  \bibinfo {pages} {240} (\bibinfo {year} {1980})}\BibitemShut {NoStop}%
\bibitem [{\citenamefont {Berry}\ \emph {et~al.}(1980)\citenamefont {Berry},
  \citenamefont {Chambers}, \citenamefont {Large}, \citenamefont {Upstill},\
  and\ \citenamefont {Walmsley}}]{Berry_Chambers_1980}%
  \BibitemOpen
  \bibfield  {author} {\bibinfo {author} {\bibfnamefont {M.~V.}\ \bibnamefont
  {Berry}}, \bibinfo {author} {\bibfnamefont {R.~G.}\ \bibnamefont {Chambers}},
  \bibinfo {author} {\bibfnamefont {M.~D.}\ \bibnamefont {Large}}, \bibinfo
  {author} {\bibfnamefont {C.}~\bibnamefont {Upstill}},\ and\ \bibinfo {author}
  {\bibfnamefont {J.~C.}\ \bibnamefont {Walmsley}},\ }\bibfield  {title}
  {\bibinfo {title} {Wavefront dislocations in the {Aharonov-Bohm} effect and
  its water wave analogue},\ }\href {https://doi.org/10.1088/0143-0807/1/3/008}
  {\bibfield  {journal} {\bibinfo  {journal} {European Journal of Physics}\
  }\textbf {\bibinfo {volume} {1}},\ \bibinfo {pages} {154} (\bibinfo {year}
  {1980})}\BibitemShut {NoStop}%
\bibitem [{\citenamefont {Alford}\ and\ \citenamefont
  {Wilczek}(1989)}]{Wilczek_PhysRevLett.62.1071}%
  \BibitemOpen
  \bibfield  {author} {\bibinfo {author} {\bibfnamefont {M.~G.}\ \bibnamefont
  {Alford}}\ and\ \bibinfo {author} {\bibfnamefont {F.}~\bibnamefont
  {Wilczek}},\ }\bibfield  {title} {\bibinfo {title} {{Aharonov-Bohm}
  interaction of cosmic strings with matter},\ }\href
  {https://doi.org/10.1103/PhysRevLett.62.1071} {\bibfield  {journal} {\bibinfo
   {journal} {Phys. Rev. Lett.}\ }\textbf {\bibinfo {volume} {62}},\ \bibinfo
  {pages} {1071} (\bibinfo {year} {1989})}\BibitemShut {NoStop}%
\bibitem [{\citenamefont {Gerbert}(1989)}]{Gerbert_PhysRevD.40.1346}%
  \BibitemOpen
  \bibfield  {author} {\bibinfo {author} {\bibfnamefont {P.~d.~S.}\
  \bibnamefont {Gerbert}},\ }\bibfield  {title} {\bibinfo {title} {Fermions in
  an {Aharonov-Bohm} field and cosmic strings},\ }\href
  {https://doi.org/10.1103/PhysRevD.40.1346} {\bibfield  {journal} {\bibinfo
  {journal} {Phys. Rev. D}\ }\textbf {\bibinfo {volume} {40}},\ \bibinfo
  {pages} {1346} (\bibinfo {year} {1989})}\BibitemShut {NoStop}%
\bibitem [{\citenamefont {Hagen}(1990)}]{Hagen_PhysRevLett.64.503}%
  \BibitemOpen
  \bibfield  {author} {\bibinfo {author} {\bibfnamefont {C.~R.}\ \bibnamefont
  {Hagen}},\ }\bibfield  {title} {\bibinfo {title} {{Aharonov-Bohm} scattering
  of particles with spin},\ }\href {https://doi.org/10.1103/PhysRevLett.64.503}
  {\bibfield  {journal} {\bibinfo  {journal} {Phys. Rev. Lett.}\ }\textbf
  {\bibinfo {volume} {64}},\ \bibinfo {pages} {503} (\bibinfo {year}
  {1990})}\BibitemShut {NoStop}%
\bibitem [{\citenamefont {Hagen}\ and\ \citenamefont
  {Ramaswamy}(1990)}]{Hagen_PhysRevD.42.3524}%
  \BibitemOpen
  \bibfield  {author} {\bibinfo {author} {\bibfnamefont {C.~R.}\ \bibnamefont
  {Hagen}}\ and\ \bibinfo {author} {\bibfnamefont {S.}~\bibnamefont
  {Ramaswamy}},\ }\bibfield  {title} {\bibinfo {title} {{Aharonov-Bohm}
  scattering of massive spin-one particles},\ }\href
  {https://doi.org/10.1103/PhysRevD.42.3524} {\bibfield  {journal} {\bibinfo
  {journal} {Phys. Rev. D}\ }\textbf {\bibinfo {volume} {42}},\ \bibinfo
  {pages} {3524} (\bibinfo {year} {1990})}\BibitemShut {NoStop}%
\bibitem [{\citenamefont {Wu}\ and\ \citenamefont {Yang}(1975)}]{Wu:NA}%
  \BibitemOpen
  \bibfield  {author} {\bibinfo {author} {\bibfnamefont {T.~T.}\ \bibnamefont
  {Wu}}\ and\ \bibinfo {author} {\bibfnamefont {C.~N.}\ \bibnamefont {Yang}},\
  }\bibfield  {title} {\bibinfo {title} {Concept of nonintegrable phase factors
  and global formulation of gauge fields},\ }\href
  {https://doi.org/10.1103/PhysRevD.12.3845} {\bibfield  {journal} {\bibinfo
  {journal} {Phys. Rev. D}\ }\textbf {\bibinfo {volume} {12}},\ \bibinfo
  {pages} {3845} (\bibinfo {year} {1975})}\BibitemShut {NoStop}%
\bibitem [{\citenamefont {Horv\'athy}(1986)}]{Horvathy:NA}%
  \BibitemOpen
  \bibfield  {author} {\bibinfo {author} {\bibfnamefont {P.~A.}\ \bibnamefont
  {Horv\'athy}},\ }\bibfield  {title} {\bibinfo {title} {Non-{Abelian}
  {Aharonov-Bohm} effect},\ }\href {https://doi.org/10.1103/PhysRevD.33.407}
  {\bibfield  {journal} {\bibinfo  {journal} {Phys. Rev. D}\ }\textbf {\bibinfo
  {volume} {33}},\ \bibinfo {pages} {407} (\bibinfo {year} {1986})}\BibitemShut
  {NoStop}%
\bibitem [{\citenamefont {Bright}\ and\ \citenamefont
  {Singleton}(2015)}]{Bright:2015rsa}%
  \BibitemOpen
  \bibfield  {author} {\bibinfo {author} {\bibfnamefont {M.}~\bibnamefont
  {Bright}}\ and\ \bibinfo {author} {\bibfnamefont {D.}~\bibnamefont
  {Singleton}},\ }\bibfield  {title} {\bibinfo {title} {{Time-dependent
  non-Abelian Aharonov-Bohm effect}},\ }\href
  {https://doi.org/10.1103/PhysRevD.91.085010} {\bibfield  {journal} {\bibinfo
  {journal} {Phys. Rev. D}\ }\textbf {\bibinfo {volume} {91}},\ \bibinfo
  {pages} {085010} (\bibinfo {year} {2015})}\BibitemShut {NoStop}%
\bibitem [{\citenamefont {Hosseini~Mansoori}\ and\ \citenamefont
  {Mirza}(2016)}]{HosseiniMansoori:2016rex}%
  \BibitemOpen
  \bibfield  {author} {\bibinfo {author} {\bibfnamefont {S.~A.}\ \bibnamefont
  {Hosseini~Mansoori}}\ and\ \bibinfo {author} {\bibfnamefont {B.}~\bibnamefont
  {Mirza}},\ }\bibfield  {title} {\bibinfo {title} {{Non-Abelian
  Aharonov\textendash{}Bohm effect with the time-dependent gauge fields}},\
  }\href {https://doi.org/10.1016/j.physletb.2016.02.004} {\bibfield  {journal}
  {\bibinfo  {journal} {Phys. Lett. B}\ }\textbf {\bibinfo {volume} {755}},\
  \bibinfo {pages} {88} (\bibinfo {year} {2016})}\BibitemShut {NoStop}%
\bibitem [{\citenamefont {Goldman}\ \emph {et~al.}(2009)\citenamefont
  {Goldman}, \citenamefont {Kubasiak}, \citenamefont {Gaspard},\ and\
  \citenamefont {Lewenstein}}]{Goldman:paper1}%
  \BibitemOpen
  \bibfield  {author} {\bibinfo {author} {\bibfnamefont {N.}~\bibnamefont
  {Goldman}}, \bibinfo {author} {\bibfnamefont {A.}~\bibnamefont {Kubasiak}},
  \bibinfo {author} {\bibfnamefont {P.}~\bibnamefont {Gaspard}},\ and\ \bibinfo
  {author} {\bibfnamefont {M.}~\bibnamefont {Lewenstein}},\ }\bibfield  {title}
  {\bibinfo {title} {Ultracold atomic gases in {non-Abelian} gauge potentials:
  The case of constant {Wilson} loop},\ }\href
  {https://doi.org/10.1103/PhysRevA.79.023624} {\bibfield  {journal} {\bibinfo
  {journal} {Phys. Rev. A}\ }\textbf {\bibinfo {volume} {79}},\ \bibinfo
  {pages} {023624} (\bibinfo {year} {2009})}\BibitemShut {NoStop}%
\bibitem [{\citenamefont {Bermudez}\ \emph {et~al.}(2010)\citenamefont
  {Bermudez}, \citenamefont {Goldman}, \citenamefont {Kubasiak}, \citenamefont
  {Lewenstein},\ and\ \citenamefont {Martin-Delgado}}]{Bermudez:paper}%
  \BibitemOpen
  \bibfield  {author} {\bibinfo {author} {\bibfnamefont {A.}~\bibnamefont
  {Bermudez}}, \bibinfo {author} {\bibfnamefont {N.}~\bibnamefont {Goldman}},
  \bibinfo {author} {\bibfnamefont {A.}~\bibnamefont {Kubasiak}}, \bibinfo
  {author} {\bibfnamefont {M.}~\bibnamefont {Lewenstein}},\ and\ \bibinfo
  {author} {\bibfnamefont {M.~A.}\ \bibnamefont {Martin-Delgado}},\ }\bibfield
  {title} {\bibinfo {title} {Topological phase transitions in the non-{Abelian}
  honeycomb lattice},\ }\href {https://doi.org/10.1088/1367-2630/12/3/033041}
  {\bibfield  {journal} {\bibinfo  {journal} {New Journal of Physics}\ }\textbf
  {\bibinfo {volume} {12}},\ \bibinfo {pages} {033041} (\bibinfo {year}
  {2010})}\BibitemShut {NoStop}%
\bibitem [{\citenamefont {Gorshkov}\ \emph {et~al.}(2010)\citenamefont
  {Gorshkov}, \citenamefont {Hermele}, \citenamefont {Gurarie}, \citenamefont
  {Xu}, \citenamefont {Julienne}, \citenamefont {Ye}, \citenamefont {Zoller},
  \citenamefont {Demler}, \citenamefont {Lukin},\ and\ \citenamefont
  {Rey}}]{Gorshkov:2010Nature}%
  \BibitemOpen
  \bibfield  {author} {\bibinfo {author} {\bibfnamefont {A.}~\bibnamefont
  {Gorshkov}}, \bibinfo {author} {\bibfnamefont {M.}~\bibnamefont {Hermele}},
  \bibinfo {author} {\bibfnamefont {V.}~\bibnamefont {Gurarie}}, \bibinfo
  {author} {\bibfnamefont {C.}~\bibnamefont {Xu}}, \bibinfo {author}
  {\bibfnamefont {P.}~\bibnamefont {Julienne}}, \bibinfo {author}
  {\bibfnamefont {J.}~\bibnamefont {Ye}}, \bibinfo {author} {\bibfnamefont
  {P.}~\bibnamefont {Zoller}}, \bibinfo {author} {\bibfnamefont
  {E.}~\bibnamefont {Demler}}, \bibinfo {author} {\bibfnamefont
  {M.}~\bibnamefont {Lukin}},\ and\ \bibinfo {author} {\bibfnamefont
  {A.}~\bibnamefont {Rey}},\ }\bibfield  {title} {\bibinfo {title} {Two-orbital
  {SU(N)} magnetism with ultracold alkaline-earth atoms},\ }\href
  {https://doi.org/10.1038/NPHYS1535} {\bibfield  {journal} {\bibinfo
  {journal} {Nature Physics}\ }\textbf {\bibinfo {volume} {6}},\ \bibinfo
  {pages} {289} (\bibinfo {year} {2010})}\BibitemShut {NoStop}%
\bibitem [{\citenamefont {Barnett}\ \emph {et~al.}(2012)\citenamefont
  {Barnett}, \citenamefont {Boyd},\ and\ \citenamefont
  {Galitski}}]{Barnett:paper}%
  \BibitemOpen
  \bibfield  {author} {\bibinfo {author} {\bibfnamefont {R.}~\bibnamefont
  {Barnett}}, \bibinfo {author} {\bibfnamefont {G.~R.}\ \bibnamefont {Boyd}},\
  and\ \bibinfo {author} {\bibfnamefont {V.}~\bibnamefont {Galitski}},\
  }\bibfield  {title} {\bibinfo {title} {{SU(3) Spin-Orbit Coupling in Systems
  of Ultracold Atoms}},\ }\href
  {https://doi.org/10.1103/PhysRevLett.109.235308} {\bibfield  {journal}
  {\bibinfo  {journal} {Phys. Rev. Lett.}\ }\textbf {\bibinfo {volume} {109}},\
  \bibinfo {pages} {235308} (\bibinfo {year} {2012})}\BibitemShut {NoStop}%
\bibitem [{\citenamefont {Banerjee}\ \emph {et~al.}(2013)\citenamefont
  {Banerjee}, \citenamefont {B\"ogli}, \citenamefont {Dalmonte}, \citenamefont
  {Rico}, \citenamefont {Stebler}, \citenamefont {Wiese},\ and\ \citenamefont
  {Zoller}}]{PhysRevLett.110.125303}%
  \BibitemOpen
  \bibfield  {author} {\bibinfo {author} {\bibfnamefont {D.}~\bibnamefont
  {Banerjee}}, \bibinfo {author} {\bibfnamefont {M.}~\bibnamefont {B\"ogli}},
  \bibinfo {author} {\bibfnamefont {M.}~\bibnamefont {Dalmonte}}, \bibinfo
  {author} {\bibfnamefont {E.}~\bibnamefont {Rico}}, \bibinfo {author}
  {\bibfnamefont {P.}~\bibnamefont {Stebler}}, \bibinfo {author} {\bibfnamefont
  {U.-J.}\ \bibnamefont {Wiese}},\ and\ \bibinfo {author} {\bibfnamefont
  {P.}~\bibnamefont {Zoller}},\ }\bibfield  {title} {\bibinfo {title} {{Atomic
  Quantum Simulation of $\mathbf{U}(N)$ and $\mathrm{SU}(N)$ Non-Abelian
  Lattice Gauge Theories}},\ }\href
  {https://doi.org/10.1103/PhysRevLett.110.125303} {\bibfield  {journal}
  {\bibinfo  {journal} {Phys. Rev. Lett.}\ }\textbf {\bibinfo {volume} {110}},\
  \bibinfo {pages} {125303} (\bibinfo {year} {2013})}\BibitemShut {NoStop}%
\bibitem [{\citenamefont {Goldman}\ \emph {et~al.}(2013)\citenamefont
  {Goldman}, \citenamefont {Gerbier},\ and\ \citenamefont
  {Lewenstein}}]{Goldman:paper2}%
  \BibitemOpen
  \bibfield  {author} {\bibinfo {author} {\bibfnamefont {N.}~\bibnamefont
  {Goldman}}, \bibinfo {author} {\bibfnamefont {F.}~\bibnamefont {Gerbier}},\
  and\ \bibinfo {author} {\bibfnamefont {M.}~\bibnamefont {Lewenstein}},\
  }\bibfield  {title} {\bibinfo {title} {Realizing {non-Abelian} gauge
  potentials in optical square lattices: an application to atomic {Chern}
  insulators},\ }\href {https://doi.org/10.1088/0953-4075/46/13/134010}
  {\bibfield  {journal} {\bibinfo  {journal} {Journal of Physics B: Atomic,
  Molecular and Optical Physics}\ }\textbf {\bibinfo {volume} {46}},\ \bibinfo
  {pages} {134010} (\bibinfo {year} {2013})}\BibitemShut {NoStop}%
\bibitem [{\citenamefont {Tagliacozzo}\ \emph {et~al.}(2013)\citenamefont
  {Tagliacozzo}, \citenamefont {Celi}, \citenamefont {Orland},\ and\
  \citenamefont {Lewenstein}}]{Tagliacozzo:2012df}%
  \BibitemOpen
  \bibfield  {author} {\bibinfo {author} {\bibfnamefont {L.}~\bibnamefont
  {Tagliacozzo}}, \bibinfo {author} {\bibfnamefont {A.}~\bibnamefont {Celi}},
  \bibinfo {author} {\bibfnamefont {P.}~\bibnamefont {Orland}},\ and\ \bibinfo
  {author} {\bibfnamefont {M.}~\bibnamefont {Lewenstein}},\ }\bibfield  {title}
  {\bibinfo {title} {{Simulations of {non-Abelian} gauge theories with optical
  lattices}},\ }\href {https://doi.org/10.1038/ncomms3615} {\bibfield
  {journal} {\bibinfo  {journal} {Nature Commun.}\ }\textbf {\bibinfo {volume}
  {4}},\ \bibinfo {pages} {2615} (\bibinfo {year} {2013})}\BibitemShut
  {NoStop}%
\bibitem [{\citenamefont {Zohar}\ \emph {et~al.}(2013)\citenamefont {Zohar},
  \citenamefont {Cirac},\ and\ \citenamefont
  {Reznik}}]{PhysRevLett.110.125304}%
  \BibitemOpen
  \bibfield  {author} {\bibinfo {author} {\bibfnamefont {E.}~\bibnamefont
  {Zohar}}, \bibinfo {author} {\bibfnamefont {J.~I.}\ \bibnamefont {Cirac}},\
  and\ \bibinfo {author} {\bibfnamefont {B.}~\bibnamefont {Reznik}},\
  }\bibfield  {title} {\bibinfo {title} {{Cold-Atom Quantum Simulator for
  {SU(2)} Yang-Mills Lattice Gauge Theory}},\ }\href
  {https://doi.org/10.1103/PhysRevLett.110.125304} {\bibfield  {journal}
  {\bibinfo  {journal} {Phys. Rev. Lett.}\ }\textbf {\bibinfo {volume} {110}},\
  \bibinfo {pages} {125304} (\bibinfo {year} {2013})}\BibitemShut {NoStop}%
\bibitem [{\citenamefont {Osterloh}\ \emph {et~al.}(2005)\citenamefont
  {Osterloh}, \citenamefont {Baig}, \citenamefont {Santos}, \citenamefont
  {Zoller},\ and\ \citenamefont {Lewenstein}}]{Osterloh:paper}%
  \BibitemOpen
  \bibfield  {author} {\bibinfo {author} {\bibfnamefont {K.}~\bibnamefont
  {Osterloh}}, \bibinfo {author} {\bibfnamefont {M.}~\bibnamefont {Baig}},
  \bibinfo {author} {\bibfnamefont {L.}~\bibnamefont {Santos}}, \bibinfo
  {author} {\bibfnamefont {P.}~\bibnamefont {Zoller}},\ and\ \bibinfo {author}
  {\bibfnamefont {M.}~\bibnamefont {Lewenstein}},\ }\bibfield  {title}
  {\bibinfo {title} {{Cold Atoms in Non-Abelian Gauge Potentials: From the
  {Hofstadter} "Moth" to Lattice Gauge Theory}},\ }\href
  {https://doi.org/10.1103/PhysRevLett.95.010403} {\bibfield  {journal}
  {\bibinfo  {journal} {Phys. Rev. Lett.}\ }\textbf {\bibinfo {volume} {95}},\
  \bibinfo {pages} {010403} (\bibinfo {year} {2005})}\BibitemShut {NoStop}%
\bibitem [{\citenamefont {Jacob}\ \emph {et~al.}(2007)\citenamefont {Jacob},
  \citenamefont {Ohberg}, \citenamefont {Juzeliunas},\ and\ \citenamefont
  {Santos}}]{Jacob:2007jb}%
  \BibitemOpen
  \bibfield  {author} {\bibinfo {author} {\bibfnamefont {A.}~\bibnamefont
  {Jacob}}, \bibinfo {author} {\bibfnamefont {P.}~\bibnamefont {Ohberg}},
  \bibinfo {author} {\bibfnamefont {G.}~\bibnamefont {Juzeliunas}},\ and\
  \bibinfo {author} {\bibfnamefont {L.}~\bibnamefont {Santos}},\ }\bibfield
  {title} {\bibinfo {title} {{Cold atom dynamics in {non-Abelian} gauge
  fields}},\ }\href {https://doi.org/10.1007/s00340-007-2865-6} {\bibfield
  {journal} {\bibinfo  {journal} {Appl. Phys. B}\ }\textbf {\bibinfo {volume}
  {89}},\ \bibinfo {pages} {439} (\bibinfo {year} {2007})}\BibitemShut
  {NoStop}%
\bibitem [{\citenamefont {Dalibard}\ \emph {et~al.}(2011)\citenamefont
  {Dalibard}, \citenamefont {Gerbier}, \citenamefont {Juzeliunas},\ and\
  \citenamefont {Ohberg}}]{Dalibard:2010ph}%
  \BibitemOpen
  \bibfield  {author} {\bibinfo {author} {\bibfnamefont {J.}~\bibnamefont
  {Dalibard}}, \bibinfo {author} {\bibfnamefont {F.}~\bibnamefont {Gerbier}},
  \bibinfo {author} {\bibfnamefont {G.}~\bibnamefont {Juzeliunas}},\ and\
  \bibinfo {author} {\bibfnamefont {P.}~\bibnamefont {Ohberg}},\ }\bibfield
  {title} {\bibinfo {title} {{{Colloquium: Artificial gauge potentials for
  neutral atoms}}},\ }\href {https://doi.org/10.1103/RevModPhys.83.1523}
  {\bibfield  {journal} {\bibinfo  {journal} {Rev. Mod. Phys.}\ }\textbf
  {\bibinfo {volume} {83}},\ \bibinfo {pages} {1523} (\bibinfo {year}
  {2011})}\BibitemShut {NoStop}%
\bibitem [{\citenamefont {Goldman}\ \emph {et~al.}(2014)\citenamefont
  {Goldman}, \citenamefont {Juzeliūnas}, \citenamefont {Öhberg},\ and\
  \citenamefont {Spielman}}]{Goldman_2014}%
  \BibitemOpen
  \bibfield  {author} {\bibinfo {author} {\bibfnamefont {N.}~\bibnamefont
  {Goldman}}, \bibinfo {author} {\bibfnamefont {G.}~\bibnamefont
  {Juzeliūnas}}, \bibinfo {author} {\bibfnamefont {P.}~\bibnamefont
  {Öhberg}},\ and\ \bibinfo {author} {\bibfnamefont {I.~B.}\ \bibnamefont
  {Spielman}},\ }\bibfield  {title} {\bibinfo {title} {Light-induced gauge
  fields for ultracold atoms},\ }\href
  {https://doi.org/10.1088/0034-4885/77/12/126401} {\bibfield  {journal}
  {\bibinfo  {journal} {Reports on Progress in Physics}\ }\textbf {\bibinfo
  {volume} {77}},\ \bibinfo {pages} {126401} (\bibinfo {year}
  {2014})}\BibitemShut {NoStop}%
\bibitem [{\citenamefont {Huo}\ \emph {et~al.}(2014)\citenamefont {Huo},
  \citenamefont {Nie}, \citenamefont {Hutchinson},\ and\ \citenamefont
  {Kwek}}]{Huo:paper}%
  \BibitemOpen
  \bibfield  {author} {\bibinfo {author} {\bibfnamefont {M.-X.}\ \bibnamefont
  {Huo}}, \bibinfo {author} {\bibfnamefont {W.}~\bibnamefont {Nie}}, \bibinfo
  {author} {\bibfnamefont {D.~A.~W.}\ \bibnamefont {Hutchinson}},\ and\
  \bibinfo {author} {\bibfnamefont {L.~C.}\ \bibnamefont {Kwek}},\ }\bibfield
  {title} {\bibinfo {title} {A solenoidal synthetic field and the {non-Abelian}
  {Aharonov-Bohm} effects in neutral atoms},\ }\href
  {https://doi.org/https://doi.org/10.1038/srep05992} {\bibfield  {journal}
  {\bibinfo  {journal} {Scientific Reports}\ }\textbf {\bibinfo {volume} {4}},\
  \bibinfo {pages} {5992} (\bibinfo {year} {2014})}\BibitemShut {NoStop}%
\bibitem [{\citenamefont {Aidelsburger}\ \emph {et~al.}(2018)\citenamefont
  {Aidelsburger}, \citenamefont {Nascimbene},\ and\ \citenamefont
  {Goldman}}]{AIDELSBURGER2018394}%
  \BibitemOpen
  \bibfield  {author} {\bibinfo {author} {\bibfnamefont {M.}~\bibnamefont
  {Aidelsburger}}, \bibinfo {author} {\bibfnamefont {S.}~\bibnamefont
  {Nascimbene}},\ and\ \bibinfo {author} {\bibfnamefont {N.}~\bibnamefont
  {Goldman}},\ }\bibfield  {title} {\bibinfo {title} {Artificial gauge fields
  in materials and engineered systems},\ }\href
  {https://doi.org/https://doi.org/10.1016/j.crhy.2018.03.002} {\bibfield
  {journal} {\bibinfo  {journal} {Comptes Rendus Physique}\ }\textbf {\bibinfo
  {volume} {19}},\ \bibinfo {pages} {394} (\bibinfo {year} {2018})}\BibitemShut
  {NoStop}%
\bibitem [{\citenamefont {Yang}\ \emph {et~al.}(2019)\citenamefont {Yang},
  \citenamefont {Peng}, \citenamefont {Zhu}, \citenamefont {Buljan},
  \citenamefont {Joannopoulos}, \citenamefont {Zhen},\ and\ \citenamefont
  {Soljačić}}]{Yang:2019Science}%
  \BibitemOpen
  \bibfield  {author} {\bibinfo {author} {\bibfnamefont {Y.}~\bibnamefont
  {Yang}}, \bibinfo {author} {\bibfnamefont {C.}~\bibnamefont {Peng}}, \bibinfo
  {author} {\bibfnamefont {D.}~\bibnamefont {Zhu}}, \bibinfo {author}
  {\bibfnamefont {H.}~\bibnamefont {Buljan}}, \bibinfo {author} {\bibfnamefont
  {J.~D.}\ \bibnamefont {Joannopoulos}}, \bibinfo {author} {\bibfnamefont
  {B.}~\bibnamefont {Zhen}},\ and\ \bibinfo {author} {\bibfnamefont
  {M.}~\bibnamefont {Soljačić}},\ }\bibfield  {title} {\bibinfo {title}
  {Synthesis and observation of {non-Abelian} gauge fields in real space},\
  }\href {https://doi.org/10.1126/science.aay3183} {\bibfield  {journal}
  {\bibinfo  {journal} {Science}\ }\textbf {\bibinfo {volume} {365}},\ \bibinfo
  {pages} {1021} (\bibinfo {year} {2019})}\BibitemShut {NoStop}%
\bibitem [{\citenamefont {N\'emeth}\ and\ \citenamefont
  {Cserti}(2023)}]{Nemeth-Cserti:paper}%
  \BibitemOpen
  \bibfield  {author} {\bibinfo {author} {\bibfnamefont {R.}~\bibnamefont
  {N\'emeth}}\ and\ \bibinfo {author} {\bibfnamefont {J.}~\bibnamefont
  {Cserti}},\ }\bibfield  {title} {\bibinfo {title} {Unified description of the
  {Aharonov-Bohm} effect in isotropic multiband electronic systems},\ }\href
  {https://doi.org/10.1103/PhysRevResearch.5.023154} {\bibfield  {journal}
  {\bibinfo  {journal} {Phys. Rev. Res.}\ }\textbf {\bibinfo {volume} {5}},\
  \bibinfo {pages} {023154} (\bibinfo {year} {2023})}\BibitemShut {NoStop}%
\bibitem [{\citenamefont {Heisenberg}(1932)}]{Heisenberg:1932}%
  \BibitemOpen
  \bibfield  {author} {\bibinfo {author} {\bibfnamefont {W.}~\bibnamefont
  {Heisenberg}},\ }\bibfield  {title} {\bibinfo {title} {{Über den Bau der
  Atomkerne. I.}},\ }\href {https://doi.org/10.1007/BF01342433} {\bibfield
  {journal} {\bibinfo  {journal} {Z. Physik}\ }\textbf {\bibinfo {volume}
  {77}},\ \bibinfo {pages} {1} (\bibinfo {year} {1932})}\BibitemShut {NoStop}%
\bibitem [{\citenamefont {Cheng}\ and\ \citenamefont
  {Li}(1984)}]{Cheng:1984vwu}%
  \BibitemOpen
  \bibfield  {author} {\bibinfo {author} {\bibfnamefont {T.-P.}\ \bibnamefont
  {Cheng}}\ and\ \bibinfo {author} {\bibfnamefont {L.-F.}\ \bibnamefont {Li}},\
  }\href@noop {} {\emph {\bibinfo {title} {{Gauge Theory of Elementary Particle
  Physics}}}}\ (\bibinfo  {publisher} {Oxford University Press},\ \bibinfo
  {year} {1984})\BibitemShut {NoStop}%
\bibitem [{\citenamefont {Nachtmann}\ \emph {et~al.}(2012)\citenamefont
  {Nachtmann}, \citenamefont {Lahee},\ and\ \citenamefont
  {Wetzel}}]{Nachtmann:1990ta}%
  \BibitemOpen
  \bibfield  {author} {\bibinfo {author} {\bibfnamefont {O.}~\bibnamefont
  {Nachtmann}}, \bibinfo {author} {\bibfnamefont {A.}~\bibnamefont {Lahee}},\
  and\ \bibinfo {author} {\bibfnamefont {W.}~\bibnamefont {Wetzel}},\
  }\href@noop {} {\emph {\bibinfo {title} {Elementary Particle Physics:
  Concepts and Phenomena}}}\ (\bibinfo  {publisher} {Springer Berlin
  Heidelberg},\ \bibinfo {year} {2012})\BibitemShut {NoStop}%
\bibitem [{\citenamefont {Horv\'ath}\ and\ \citenamefont
  {Tr\'ocs\'anyi}(2019)}]{Horvath:2019atn}%
  \BibitemOpen
  \bibfield  {author} {\bibinfo {author} {\bibfnamefont {D.}~\bibnamefont
  {Horv\'ath}}\ and\ \bibinfo {author} {\bibfnamefont {Z.}~\bibnamefont
  {Tr\'ocs\'anyi}},\ }\href@noop {} {\emph {\bibinfo {title} {{Introduction to
  Particle Physics}}}}\ (\bibinfo  {publisher} {Cambridge Scholars
  Publishing},\ \bibinfo {year} {2019})\BibitemShut {NoStop}%
\bibitem [{\citenamefont {Erd{\'e}lyi}\ \emph {et~al.}(1953)\citenamefont
  {Erd{\'e}lyi}, \citenamefont {Magnus}, \citenamefont {Oberhettinger},\ and\
  \citenamefont {Tricomi}}]{HTF:book}%
  \BibitemOpen
  \bibfield  {author} {\bibinfo {author} {\bibfnamefont {A.}~\bibnamefont
  {Erd{\'e}lyi}}, \bibinfo {author} {\bibfnamefont {W.}~\bibnamefont {Magnus}},
  \bibinfo {author} {\bibfnamefont {F.}~\bibnamefont {Oberhettinger}},\ and\
  \bibinfo {author} {\bibfnamefont {F.~G.}\ \bibnamefont {Tricomi}},\
  }\href@noop {} {\emph {\bibinfo {title} {Higher Transcendental Functions.
  {V}ol. {II}}}}\ (\bibinfo  {publisher} {McGraw-Hill Book Company, Inc.},\
  \bibinfo {address} {New York-Toronto-London},\ \bibinfo {year}
  {1953})\BibitemShut {NoStop}%
\bibitem [{\citenamefont {Gell-Mann}(1964)}]{GELLMANN1964214}%
  \BibitemOpen
  \bibfield  {author} {\bibinfo {author} {\bibfnamefont {M.}~\bibnamefont
  {Gell-Mann}},\ }\bibfield  {title} {\bibinfo {title} {A schematic model of
  baryons and mesons},\ }\href
  {https://doi.org/https://doi.org/10.1016/S0031-9163(64)92001-3} {\bibfield
  {journal} {\bibinfo  {journal} {Physics Letters}\ }\textbf {\bibinfo {volume}
  {8}},\ \bibinfo {pages} {214} (\bibinfo {year} {1964})}\BibitemShut {NoStop}%
\bibitem [{\citenamefont {Zweig}(1964)}]{Zweig:1964jf}%
  \BibitemOpen
  \bibfield  {author} {\bibinfo {author} {\bibfnamefont {G.}~\bibnamefont
  {Zweig}},\ }\href@noop {} {\emph {\bibinfo {title} {{An SU$_3$ model for
  strong interaction symmetry and its breaking; Version 1}}}},\ \bibinfo {type}
  {Tech. Rep.}\ (\bibinfo  {institution} {CERN},\ \bibinfo {address} {Geneva},\
  \bibinfo {year} {1964})\ \bibinfo {note} {an updated version of the paper is
  available as CERN-TH-412, 1964.}\BibitemShut {Stop}%
\end{thebibliography}
\end{document}